# Direct Visualization of a Disorder Driven Electronic Smectic Phase in Nonsymmorphic Square-Net Semimetal GdSbTe

Balaji Venkatesan,[1,2,3]§ Syu-You Guan[3]§, Jen-Te Chang[3], Shiang-Bin Chiu[1], Po-Yuan Yang[4], Chih-Chuan Su[3], Tay-Rong Chang[4], Kalaivanan Raju[3], Raman Sankar[3], Somboon Fongchaiya[5,6], Ming-Wen Chu[5], Chia-Seng Chang[1,2,3], Guoqing Chang[7], Hsin Lin[3], Adrian Del Maestro[8,9,10], Ying-Jer Kao[1,11]*, Tien-Ming Chuang[3]*

[1] Department of Physics, National Taiwan University, Taipei 10617, Taiwan

[2] Nano Science and Technology Program, Taiwan International Graduate Program, Academia Sinica and National Taiwan University, Taipei 11529, Taiwan

[3] Institute of Physics, Academia Sinica, Taipei 11529, Taiwan

[4] Department of Physics, National Cheng-Kung University, Tainan 701, Taiwan

[5] Center for Condensed Matter Sciences, National Taiwan University, Taipei 10617, Taiwan

[6] Molecular Science and Technology Program, Taiwan International Graduate Program, Academia Sinica, Taipei 11529, Taiwan

[7] Division of Physics and Applied Physics, School of Physical and Mathematical Sciences, Nanyang Technological University, 21 Nanyang Link 637371, Singapore

[8] Department of Physics and Astronomy, University of Tennessee, Knoxville, TN 37996, USA

[9] Institute for Advanced Materials & Manufacturing, University of Tennessee, Knoxville, TN 37996, USA

[10] Min H. Kao Department of Electrical Engineering and Computer Science, University of Tennessee, Knoxville, TN 37996, USA

[11] Center for Quantum Science and Technology, National Taiwan University, Taipei 10607, Taiwan

§ These authors contributed equally to this work.

*e-mail: yjkao@phys.ntu.edu.tw; chuangtm@gate.sinica.edu.tw

## Abstract

Electronic liquid crystal (ELC) phases are spontaneous symmetry breaking states believed to arise from strong electron correlation in quantum materials such as cuprates and iron pnictides. Here, we report a direct observation of a smectic phase in a weakly correlated nonsymmorphic square-net semimetal $GdSb_xTe_{2-x}$. Incommensurate smectic charge modulation and intense local unidirectional nanostructure, which coexist with Dirac fermions across Fermi level, are visualized by using spectroscopic imaging - scanning tunneling microscopy. As materials with highly mobile carriers are mostly weakly correlated, the discovery of such an ELC phase are anomalous and raise questions on the origin of their emergence. Specifically, we demonstrate how chemical substitution generates these symmetry breaking phases before the system undergoes a charge density wave (CDW) - orthorhombic structural transition. Our results highlight the importance of impurities in realizing ELC phases and present a new material platform for exploring the interplay among quenched disorder, Dirac fermions and electron correlation.

## Introduction

Electronic liquid crystal (ELC) states are enigmatic characteristic broken symmetry states in strongly correlated electronic systems[1]. Electrons localized by strong Coulomb repulsion become mobile when charge disorders are introduced into the system, leading to ELC states. A nematic phase breaks rotational symmetry while a smectic (1D stripe or 2D checkerboard) phase breaks both rotational and translational symmetry. In cuprates, where the parent state is a Mott insulating antiferromagnet, both smectic and nematic phases occur in the pseudogap regime upon hole doping the half-filled *d*-orbtials in the $CuO_2$ plane[2]. Similarly, nematicity in iron pnictide superconductors (FeSCs) occurs as either electron or hole doping of the *d*-orbitals in the Fe-pnictogen plane with spin density wave order[3]. These observed ELC phases are consistent with an interpretation that strong electronic correlation is an essential ingredient in their realization. Majority of the currently known topological materials are non-interacting or only weakly correlated due to large kinetic energy of Dirac/Weyl fermions compared to on-site Coulomb interactions. Thus, the rarity of ELC states observed in topological systems had supported the belief that they are a result of correlations.

ELC phases typically arise in materials with *d*-orbital electrons due to the localized nature of electron correlation. While ELC phases are also theorized to exist in interacting *p*-orbital optical lattices, they have yet been experimentally observed in real materials[4,5]. *Ln*$Sb_xTe_{2-x}$ (*Ln* = lanthanide, Fig. 1**a**), which is isostructural to the prototypical Dirac nodal line (DNL) semimetal ZrSiS[6], consists of bands dominated by half-filled 5*p*-orbitals within the Sb square-net layer near Fermi level, $E_F$ (Fig. 1**b**)[7], offering an ideal platform for searching these symmetry-breaking states in a 2D *p*-orbital lattice. Previous theoretical studies have established that a 2D Sb square-net can be stabilized when the structure is hypervalently bonded with six electrons per Sb atom[8,9]. Electron doping into the Sb square-net increases the Peierls instability and consequently drives *Ln*$Sb_xTe_{2-x}$ to undergo a tetragonal-to-orthorhombic/CDW phase transition[10-13]. Such a CDW phase was found to gap out topologically trivial bands near $E_F$, yielding a clean DNL band structure in $GdSb_{0.46}Te_{1.48}$[13]. Thus, *Ln*$Sb_xTe_{2-x}$, where DNL, magnetism and CDW order co-exist[10-17], also provides a fertile ground for the exploration of emergent quantum phenomena.

Here, we report a surprising discovery of a smectic phase in weakly correlated $GdSb_xTe_{2-x}$ by using spectroscopic imaging – scanning tunneling microscopy (SI-STM). Our STM images reveal an incommensurate smectic charge modulation with a periodicity of ~12.3a (a = 4.3 Å , the lattice constant of Te-Te) and intense local unidirectional nanostructures in $GdSb_{0.87}Te_{1.11}$, which is reminiscent of the "checkerboard" pattern observed in underdoped cuprates, before the CDW-structural transition. We further demonstrate that the half-filled 5*p*-orbitals in the Sb-layers, when coupled with nonsymmorphic symmetry, not only exhibits the linear band crossing but also is susceptible to the symmetry-breaking driven by disorder rather than strong electron-electron interaction, highlighting the importance of impurities in the formation of ELC phases. Taken together, this opens a pathway towards further microscopic understanding of ELC phases in other quantum materials, such as cuprates and FeSCs.

## Results

### Spectroscopic imaging of $GdSb_xTe_{2-x}$

$GdSb_xTe_{2-x}$ crystallizes in the tetragonal structure (space group *P4/nmm*) and evolves to the orthorhombic CDW phase (space group *Pmmn*) when $x < 0.80$[11,12]. $GdSb_xTe_{2-x}$ becomes antiferromagnetic below its Néel temperature, $T_N$ ~ 12 K and exhibits additional magnetic transitions in the orthorhombic phase. Slab calculation of tetragonal GdSbTe shows a similar band structure as ZrSiS, where the most prominent features are the linear band crossing between $\overline{\Gamma \mathrm{M}}$ and the surface state along $\overline{\mathrm{XM}}$, dominated by Sb 5*p*-orbitals (Fig. 1**c** and details in Fig. S1 in Supporting Information (SI)). For the STM study, we cleave $GdSb_xTe_{2-x}$ single crystals at T < 20K in ultrahigh vacuum (UHV) and then insert the samples into the STM head of our homemade UHV SI-STM system[18]. The cleavage reveals the Te-termination, yielding a charge neutral and large atomically flat surface. We first show topographic images, $z(\boldsymbol{r}, E = -1\ \mathrm{eV})$ of $GdSb_xTe_{2-x}$ (x = 0.98, 0.87 and 0.67) for comparison (Fig. 1**d**~1**f** and details in Fig. S2). These images reveal the effect of increasing Te-substitution into the Sb square-net layer ($Te_{Sb}$-substitution for brevity hereafter). STM images show the tetragonal lattice for x = 0.98 and 0.87 and an incommensurate CDW with a periodicity ~ 6.5a are observed for x = 0.67, consistent with the bulk structure phase diagram[11,12]. Electronic inhomogeneity increases with more $Te_{Sb}$-subsitution, apparent from the bright and dark areas in topographic images.

We focus our quasiparticle scattering interference (QPI) imaging measurements on tetragonal $GdSb_xTe_{2-x}$ (x = 0.87 and 0.98). Spatial LDOS modulations propagate with energy can be observed in the normalized differential conductance maps, $L(\boldsymbol{r}, E) \equiv dI/dV(\boldsymbol{r}, E)/(I(\boldsymbol{r}, E)/V)$ at T = 4.2 K (Fig. S3). Corresponding Fourier transformed images, $L(\boldsymbol{q}, E)$, reveal dispersing $\boldsymbol{q}$-vectors as shown at selective energies E = 0 and 0.5 eV for both samples (Fig. 2**a** ~ 2**d**, details in Fig. S3). We identify three linearly dispersing $\boldsymbol{q}$-vectors: $\boldsymbol{q}_1$ as the intraband scattering within the diamond-shaped band at $\overline{\Gamma}$, $\boldsymbol{q}_2$ and $\boldsymbol{q}_3$ as the interband scattering between the surface bands at $\overline{\mathrm{X}}$ (Fig. 2**e**). Our QPI results, which appears identical for both samples except with a chemical potential shift ~ 230 meV, are in excellent agreement with the joint DOS simulation (Fig. 2**f** ~ 2**h** and Fig. S4) and previous ARPES measurement[19], demonstrating the unique Dirac band structure in GdSbTe. We note that the absence of band renormalization in both QPI and ARPES measurements[19] suggests electron correlation is weak in GdSbTe. Similar Sommerfeld coefficient in ZrSiS (6.84 mJ/mol-$K^2$)[20] and GdSbTe (7.6 mJ/mol-$K^2$)[14], indicating comparable effective mass, also supports this notion.

**Incommensurate smectic charge modulation and local unidirectional nanostructure**

Interestingly, we find three non-dispersive peaks in the $L(\boldsymbol{q}$, E > -0.3 eV) images of $GdSb_{0.87}Te_{1.11}$ (Fig. 2**a**, inset) but these peaks are absent in the Fourier analysis of $z(\boldsymbol{r}$, E = -1 eV) (Fig. 1**e**). When taking the topograph of $GdSb_{0.87}Te_{1.11}$ with the bias of 1V (Fig. 3**a**), we observe not only stronger electronic inhomogeneity dominating over the lattice signal, but also four peaks of $\boldsymbol{q_S} \sim \pm 2\pi/12.3\boldsymbol{a}$ in its Fourier analysis (Fig. 3**b**). Thus, the static peaks near the Bragg peaks are the supermodulation, i.e. $\boldsymbol{q_M} = 2\pi/\boldsymbol{a} \pm \boldsymbol{q_S}$, breaking the underlying translational crystalline symmetry. The energy dependence of these peaks and the absence of the lattice superstructure from the electron diffraction (Fig 3**a**, inset) indicate they are of electronic origin. We also emphasize that this behavior differs from the conventional surface reconstruction or CDW, in which STM images show the same reconstructed structure in the filled and the empty state but with different contrast, e.g. 2H-$NbSe_2$ (Fig. S2). To better visualize these non-dispersive charge modulations, we acquire conductance maps in a larger field of view (FOV) (Fig. S5 and S6**a**). Importantly, Fourier analysis of these dI/dV maps reveals $\boldsymbol{q_S}$ breaks $C_4$-symmetry around E ≥ 0.5 eV (the inset of Fig. 3**c** and Fig. S6**b**), indicating the charge modulation is smectic.

Inverse Fourier transform of $\boldsymbol{q_S}$ from d*I/dV*($\boldsymbol{q}$, E) images yields a "checkerboard"-like pattern, $dI/dV_C$ ($\boldsymbol{r}$, E = 0.8 eV) (Fig. 3**c** and details in Fig. S5**f** - S6), which may result from the domains of 1D stripes. To analyze the tendency to form stripes along $S_1/S_2$, we calculate the local Ising-like order parameter (OP), $\Sigma(\boldsymbol{r}) = (\left|\Phi_y(\boldsymbol{r})\right|^2 - |\Phi_x(\boldsymbol{r})|^2)/(|\Phi_x(\boldsymbol{r})|^2 + \left|\Phi_y(\boldsymbol{r})\right|^2)$ from Fig. 3**c**, following two previous theoretical approaches[21,22]. Here, $\Phi_x$ and $\Phi_y$ represent the OP of CDW along *x* and *y*-axis and $\Sigma = +1$ (-1) represents the pure stripe along *y* (*x*) (details in Supplementary Note 1 and Fig. S7~S9). $\Sigma(\boldsymbol{r}$, E= 0.8 eV) map in Fig. 3**d** reveals that most regions favor the charge modulation along $S_2$ (red) over $S_1$ (blue), but the majority of $|\Sigma(\boldsymbol{r})|$ is smaller than 1, indicating the overlapping stripes. We further perform a correlation length analysis of CDW to determine whether the smectic charge modulation is checkerboard or stripe, by computing the effective Ising-like OP, $\tilde{\Sigma}(\boldsymbol{r})$ on $I(\boldsymbol{r}$, E) images (Supplementary Note 1). We obtain the averaged value of 0.56 for $\tilde{\Sigma}$, which is near the boundary between checkerboard ($\tilde{\Sigma} < 0.5$) and stripe ($\tilde{\Sigma} >$ 0.5), suggesting $GdSb_{0.87}Te_{1.11}$ is near criticality. Further investigation with systematic doping dependence is required to fully understand the impact of this putative critical point.

In addition to $\boldsymbol{q_S}$, we notice unidirectional features with a length scale of few nanometers in $z(\boldsymbol{r}$, E = 1 eV) (Fig. 3**a**) without long range order. After filtering the lattice, $\boldsymbol{q_S}$, $\boldsymbol{q_M}$ and QPI signals from a current map, $I(\boldsymbol{r}$, E = 0.8 eV), we obtain a Laplacian-enhanced image, $\nabla^2 I_N(\boldsymbol{r}$, E = 0.8 eV) (Fig. 3**e** and 3**f**), showing the existence of local unidirectional electronic nanostructure. These observations demonstrate that tetragonal $GdSb_{0.87}Te_{1.11}$ exhibits a novel ELC phase, reminiscent of underdoped cuprates[23] (comparison in Fig. S10).

**The origin of the ELC phase formation**

So far, the STM measurements are conducted in the antiferromagnetic state at T < $T_N$ ~ 12K. To elucidate whether the observed ELC phase is related to the magnetic order in $GdSb_{0.87}Te_{1.11}$, we perform STM measurements in the non-magnetic state. We find the smectic modulation persists even at T = 105 K, a temperature several times $T_N$ (Fig. S11 ~ S12) and the crystal is still tetragonal (Fig 3**a**, inset). Hence, we rule out magnetic order, spin fluctuation and structural transition as the origin for the ELC phase formation.

Alternatively, the ELC phase may be induced by chemical disorder as in the case of cuprates[2,24,25]. We note that Sb atoms naturally form a $C_2$-symmetric bonding network with neighboring Gd-Te layers due to nonsymmorphic symmetry, depending on their location within the unit cell (center or corner, Fig. 1**a**). Thus, it is conceivable that $Te_{Sb}$–substitution will create two types of anisotropic impurity states with an orientation of 90° to each other[26]. This picture is supported by our surface charge distribution calculation (Fig. 4**a** and Fig. S13). Simulated STM images with such $C_2$-symmetric impurity states randomly distributed to different Sb-sites are also in excellent agreement with our data (Fig. S14), suggesting these dense $C_2$-symmetric impurity states are responsible for the local unidirectional nanostructure and the appearance of $\boldsymbol{q_S}$. We recall that a dopant-induced anisotropic impurity state has also been observed in the nematic phase of $Ca(Fe_{1-x}Co_x)_2As_2$[27,28].

To prove this hypothesis, we examine $z(\boldsymbol{r}$, E = 1 eV) of $GdSb_{0.98}Te_{1.02}$ (Fig 4**b**) and confirm isolated Te-impurities indeed exhibit Sb-site dependent $C_2$-symmetric impurity states along the Te-Te direction (Fig. 4**c**, Fig. S13 and Fig. S15). Together with the absence of $\boldsymbol{q_S}$ and $\boldsymbol{q_M}$ in $GdSb_{0.98}Te_{1.02}$, these results provide the unambiguous evidence that the $Te_{Sb}$-subsitution induces both local unidirectional nanostructure and the smectic charge modulation.

**Minimal free energy model for disorder-driven electronic symmetry breaking**

The question now is how the ensemble of $C_2$-symmetric impurity states can also lead to the smectic charge modulation? Theoretically, a random field from the quench disorder can perturb the underlying unidirectional CDW order in underdoped cuprates, leading to a checkerboard or striped charge modulation[21,22]. To simulate the observed smectic phase, we construct a phenomenological model of an incommensurate CDW in the presence of quenched disorder following Del Maestro *et al.*[21] and employ our experimental parameters to describe the total free energy, $F_{total} = F_{clean} + F_{impurity}$, consistent with the symmetries of a square lattice (details in Supplementary Note 2 and Fig. S16~S18). The result is an effective 2D theory in terms of two complex order parameters $\Phi_x$ and $\Phi_y$, given by:

$$F_{clean} = \int d^2r[\left(|\partial_x\Phi_x|^2 + |\partial_y\Phi_y|^2\right) + \left(|\partial_y\Phi_x|^2 + |\partial_x\Phi_y|^2\right) + s\left(|\Phi_x|^2 + |\Phi_y|^2\right) + \frac{|u|}{2}\left(|\Phi_x|^2 + |\Phi_y|^2\right)^2 + \nu|\Phi_x|^2|\Phi_y|^2] \quad (1)$$

where $\Phi_x$ and $\Phi_y$ capture a local modulation of charge density defined as:

$$\delta\rho = Re\left[\Phi_x e^{iK_x\cdot r}\right] + Re[\Phi_y e^{iK_y\cdot r}] \quad (2)$$

A uniform charge density corresponds to temperature-dependent coefficient $s > 0$, whereas $s < 0$ for broken symmetry phases that can be either a checkerboard ($\nu < 0$), or striped phase ($\nu > 0$) (Fig. S18). To simulate the disorder effect, we include free energy of impurities, which is given as the following:

$$F_{impurity} = -\int d^2r\left(H_x^*\Phi_x + H_y^*\Phi_y + c.c.\right) \quad (3)$$

where $H_x$ and $H_y$ are two identical independent complex random fields (Supplementary Note 2). Fig. 4**d** shows the spatial density distribution resulting from the numerical minimization of total free energy, consistent with the experimental smectic pattern in Fig. 3**c**. It further supports that the $Te_{Sb}$-subsitution can prompt a smectic phase from a nearby CDW ground state. In contrast, the CDW order in *Ln*$Te_3$ are suppressed by intercalation[29] or isoelectronic substitution[30]. Therefore, our results establish that $Te_{Sb}$- subsitution induces both smectic charge modulation and local unidirectional nanostructure in tetragonal GdSbTe.

**Discussion**

One question regarding our SI-STM observation is whether the ELC state exists only on the surface of $GdSb_xTe_{2-x}$? Since dopant induced anisotropic impurity states should exist in the bulk and Peierls instability should take place in Sb square-net layers throughout the crystal, we deduce that the smectic charge modulation is a bulk property that could be detected by x-ray scattering or other techniques. A recent report of bond density wave order in the CDW phase of CeSbTe also signifies the importance of bonding between 5*p*-orbitals in *Ln*SbTe[31]. Thus, the spectroscopic features of Peierls instability due to the increasing electron count in the Sb square net and its relation with the smectic /CDW phase in GdSbTe requires further investigations by optical spectroscopy[12,32] or inelastic tunneling spectroscopy[33]. Its impact on in-plane electrical transport likely depends on the spatial distribution of stripe domains (e.g. Fig 3**d**) and the scattering rates of the anisotropic impurity states[28]. One may attribute the observed broken symmetry to the structural disorder from inhomogeneous spatial distribution of $Te_{Sb}$. Additionally, some form of short range CDW due to a self-interfering standing wave of Friedel oscillations is possible[34]. However, the long wavelength of smectic modulation and the $C_4$-symmetric lattice exist at the same temperature (T ~ 105 K). Thus, our results do not support the previous conjecture that the formation of ELC phases requires strong electron correlation. Rather, it underscores the essential role of charge disorder in generating the ELC phase in weakly correlated $GdSb_xTe_{2-x}$.

To further highlight the uniqueness of our results in a weakly correlated 5*p*-orbital square-net system, we compare the quenched disorder effect in the cases of *d*-orbital square-net with strong correlation. In cuprates, doping creates $C_{2v}$-symmetric pseudogap nanoclusters which are embedded in the $C_4$-symmetric insulator and percolate with increasing doping level[25]. A recent study also demonstrated the pseudogap phase favors the stripe order over the checkerboard throughout the superconducting doping range[35]. In FeSCs, the electronic nematic state has been universally observed, which is found to drive the structural transition, and the maximum nematic susceptibility coincides with the optimal Tc[3]. Recent observation of the electronic stripe phase in S-doped FeSe, which is suggested to be an impurity-pinned dynamic charge order, is in line with our observation[36]. Nematicity is also observed in correlated Dirac semimetal $BaNiS_2$ with *d*-orbital Ni square-net[37]. It will be interesting to understand how the rotational symmetry-breaking evolves as the system is doped toward the Mott insulating state with increasing electron correlation. Together with our findings, these phenomena point to the universal and important role of quenched disorder in generating and stabilizing ELC phases across a broad spectra of quantum materials with various correlation strength and less-localized carriers.

Since multiple electronic orders, magnetic and structural transition coexist in *Ln*SbTe, the interplay between charge, orbital and spin order should generate emergent electronic phases and allow for the control of their functionalities by doping or external stimuli[17,38-40]. Indeed, recent studies on NdSbTe[41], CeSbTe[10,42] and a related compound, $GdTe_3$[43] have shown that the magnetic structure is intimately linked to the CDW order, possibly via the Ruderman-Kittel-Kasuya-Yosida interaction[41,43], allowing for realizing a variety of topological phases by applying a magnetic field or by tuning the CDW order with $Te_{Sb}$-doping[17]. Both smectic charge modulation and DNL are related to 5*p*-orbitals of Sb-layers, yet they seem to coexist independently near $E_F$ in our current study. Furthermore, it appears the short range unidirectional nanostructures connect to form a ladder-like pattern at higher doping level, as observed in underdoped cuprates[25] and CeSbTe[31], with a period coincide with the CDW order (Fig. S10). How magnetism, ELC, CDW or DNL intertwine and evolve with doping in $LnSb_xTe_{2-x}$ calls for further investigations.

In summary, we find that a smectic charge modulation is induced by $Te_{Sb}$-subsitution in weakly correlated GdSbTe due to enhanced Peierls instability before the system enters the orthorhombic CDW phase. The observed smectic phase is in stark contrast from the stripes in underdoped cuprates and the predicted CDW phases in DNL semimetals[44,45], both of which require strong electron correlation. We propose a new phase diagram of $GdSb_xTe_{2-x}$ in Fig. 4**e** based on our findings. Our observation of the disorder driven smectic phase along with magnetic order and various correlation strength among different *Ln*SbTe (Table S1) represents a highly tunable DNL system that can lead to unique electronic phases. Our overall findings advance the microscopic understanding of ELC phases in strongly correlated electron systems and may ultimately allow for improved control of phase diagrams in quantum materials.

## Methods

**Single crystal growth and characterizations:** GdSbTe single crystals were grown from chemical vapor transport (CVT) [11] and flux growth techniques. Initially the polycrystalline powders of GdSbTe were prepared from the solid state synthesis process. The stoichiometric amount of high pure Gd pieces (99.999%), Sb slug (99.999%) and Te slug (99.999%) were taken into the carbon coated silica ampule and sealed in high vacuum to avoid the oxidation of the samples. The mixed compounds were heated at 700 ℃ and 900 ℃ for 24 hours and followed with intermediate grinding within the argon filled glove box. For the CVT process, the synthesized polycrystalline powders were mixed along with 200 mg of iodine and sealed by the quartz ampule with length of 40 cm with an inside pressure of about $10^{-3}$ Torr. The sealed quartz ampule was kept at a horizontal two zone furnace. The charge end and the growth end of the ampule were positioned at the hot and the cold zone of the furnace with a constant temperature of 1000 ℃ and 900 ℃ for 200 hours. After the growth was completed, the temperature of the furnace was cooled down to room temperature at a rate of 2 ℃ / min. The cold end of the ampule provided the opportunity to harvest high quality single crystals used in this study.

CVT crystal growth techniques produce a stoichiometric imbalance in GdSbTe[11]. Thus, we prepared single crystals of nearly exact stoichiometry by the self-flux method. High pure Gd pieces (99.999%), Sb slug (99.999%) and Te slug (99.999%) with the 1:30:1 molar ratios were taken into the carbon coated silica ampule and sealed with inside pressure of ~ $10^{-3}$ Torr. The ampule was kept in to the crucible furnace with the constant heating of 1050 ℃ over 30 hours for the homogeneous melt solution and further the temperature was slowly cool down to 850 ℃ with 2 ℃ /hr. The excess amount of Sb flux was removed by centrifuging at the same temperature. To avoid surface oxidation, the high quality $GdSb_{0.98}Te_{1.02}$ single crystals were collected from an argon filled grove box. The chemical composition of our crystals were measured by using an electron probe micro-analyzer (EPMA) in a scanning electron microscope. Temperature dependent magnetic susceptibility was measured by using Quantum Design MPMS magnetometer.

**STM**: To minimize the uniaxial strain on the sample during cool down, the crystal was positioned as close as possible to the geometric center of a titanium sample holder and glued with silver epoxy (EPO-TEK H20E). This ensures the thermal stress is symmetric during cool down. A titanium rod was glued to the top surface of the crystal for in situ cleavage. The matching thermal expansion coefficients of the sample holder and the rod minimizes the thermal stress on the sample during cooling. Chemically etched tungsten wires were used for STM tips. A standard AC lock-in technique was used for differential conductance measurements. The experimental conditions and parameters are specified in the captions of STM images.

**TEM:** The selected-area electron diffraction study was conducted on a FEI Tecnai, operated at 200 kV. The specimen was prepared by mechanical polishing, followed by ion milling. A liquid-nitrogen specimen holder was then exploited for the low-temperature experiments at T = 100 K.

**Density functional theory (DFT):** First-principles calculations utilizing DFT were conducted employing the Vienna ab Initio Simulation (VASP) Package[46] in the framework of the projector augmented wave method. The Perdew-Burke-Ernzerhof (PBE)[47] exchange-correlation functional was applied, with self-consistent inclusion of SOC in calculations employing a Monkhorst-Pack 17 × 17 × 1 *k*-point mesh. To account for the electron-electron interactions of 4*f* states of Gd, the on-site Coulomb repulsion energy U was considered within the GGA + U scheme[48], employing an effective $U_{eff} = (U - J) = 6$ eV. All the calculations were executed using experimental lattice constants. The surface electronic structure computation was performed with a slab model of 3-unit-cell thickness; a vacuum region with thickness larger than 15 Å. For the charge density simulation, a supercell of 5 × 5 in the in-plane directions was utilized, employing a 5 × 5 × 1 *k*-point mesh for the charge density calculation.

## Data availability

The datasets generated during the current study are available from the corresponding author on reasonable request.

**Acknowledgments:** We are indebted to Yoshiyuki Iizuka for his assistance in EPMA measurements. We are grateful to Guang-Yu Guo, J.C. Séamus Davis, Eun-Ah Kim, Ying-Ting Hsu, Tetsuo Hanaguri, Christopher Butler, Peter Wahl, Kazuhiro Fujita, Mark H. Fischer, Milan P. Allan, Ting-Kuo Lee, Sungkit Yip, Cheng-Yu Huang, Chung-Ting Ke, Chen-Hsuan Hsu, Yu-Chieh Wen, Pengcheng Dai and Yuhki Kohsaka for the helpful discussions. This work is supported by National Science and Technology Council (NSTC) in Taiwan under Grant No. NSTC 112-2112-M-001-046-MY3 (T.M.C), NSTC 110-2112-M-002-034-MY3 (Y.J.K), NSTC 111-2124-M-001-007 (R.S.), NSTC-110-2112-M-001-065-MY3 (R.S.), NSTC 111-2112-M-001-024-MY2 (S.Y.G.), partially by NSTC 112-2124-M-A49-003 (R.S., T.M.C.), by Academia Sinica under Grant No. AS-iMATE-113-12 (T.M.C., R.S., M.W.C), AS-iMATE-113-15 (H.L.) and by National Taiwan University under Grant No. NTU-CC-113L891601 (Y.J.K.). T.-R.C. was supported by the 2030 Cross-Generation Young Scholars Program from NSTC (MOST 111-2628-M-006-003-MY3), National Cheng Kung University (NCKU), Taiwan, and National Center for Theoretical Sciences, Taiwan. This research was supported, in part by Higher Education Sprout Project, Ministry of Education to the Headquarters of University Advancement at NCKU. A.D. acknowledges partial support by the National Science Foundation Materials Research Science and Engineering Center program through the UT Knoxville Center for Advanced Materials and Manufacturing (DMR-2309083). T.M.C. is grateful for the support of Golden-Jade Fellowship from Kenda Foundation.

**Author contributions:** Conceptualization: TRC, RS, MWC, HL, AD, YJK, TMC. Investigation: Crystal growth and characterizations (KR, RS), STM measurements and analysis (BV, SYG, JTC, CCS, TMC), TEM measurements and analysis (SF, MWC), DFT calculation and JDOS simulation (PYY, TRC, GC, HL) and Theoretical model/simulation for charge modulation (SBC, AD, YJK). Supervision: TRC, RS, MWC, CSC, HL, AD, YJK, TMC. Writing: All authors.

**Competing interests:** Authors declare that they have no competing interests.

## Figures

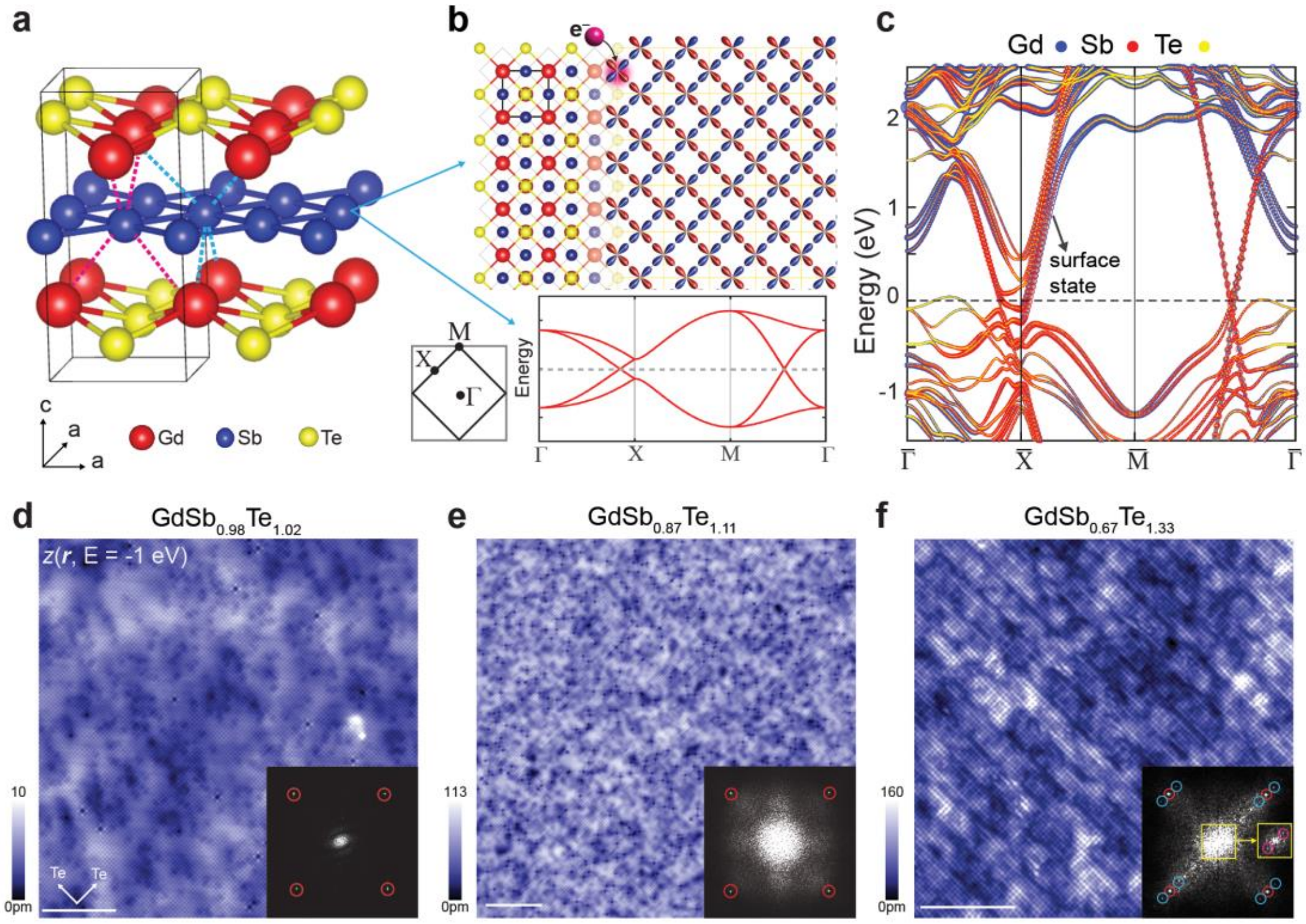

**Fig. 1. Overview of $GdSb_xTe_{2-x}$. a** Crystal structure of $GdSb_xTe_{2-x}$. The gray cuboid represents the tetragonal unit cell. Sb atoms form two anisotropic bonding networks with Gd-Te layers due to non-symmorphic symmetry. The network from the center Sb atom (red dashed line) and that from the corner Sb atom (cyan dashed line) rotate 90° with respect to each other. **b** Schematic of 5*p*-orbitals in the Sb square-net layer (top). Bulk band structure of the half-filled $p_x$ and $p_y$ orbitals of a 4 × 4 square-net system due to band folding (bottom). The black square represents the unit cell and Brillouin zone. Nodal lines at X are protected by nonsymmorphic symmetry and robust against SOC while those along ΓX and ΓM protected by mirror-symmetry are gapped by SOC. **c** Surface band structure of GdSbTe in the antiferromagnetic state from slab calculations. Orbital contribution from Gd, Sb and Te is denoted by blue, red and yellow circle, respectively. **d - f** Topographic images of $GdSb_{0.98}Te_{1.02}$, $GdSb_{0.87}Te_{1.11}$, and $GdSb_{0.67}Te_{1.33}$ measured at T = 4.2 K (V = -1 V, I = 20 pA), respectively. The arrows indicate the Te-Te direction and the scale bar represents 10 nm. The insets show the corresponding Fourier analysis, where Bragg, CDW and the supermodulation peaks are marked with red, magenta and cyan circles. The yellow square in the inset of (**f**)

depicts the enlarged area in the center with adjusted contrast to show CDW peaks. Additional comparison in Fig. S2.

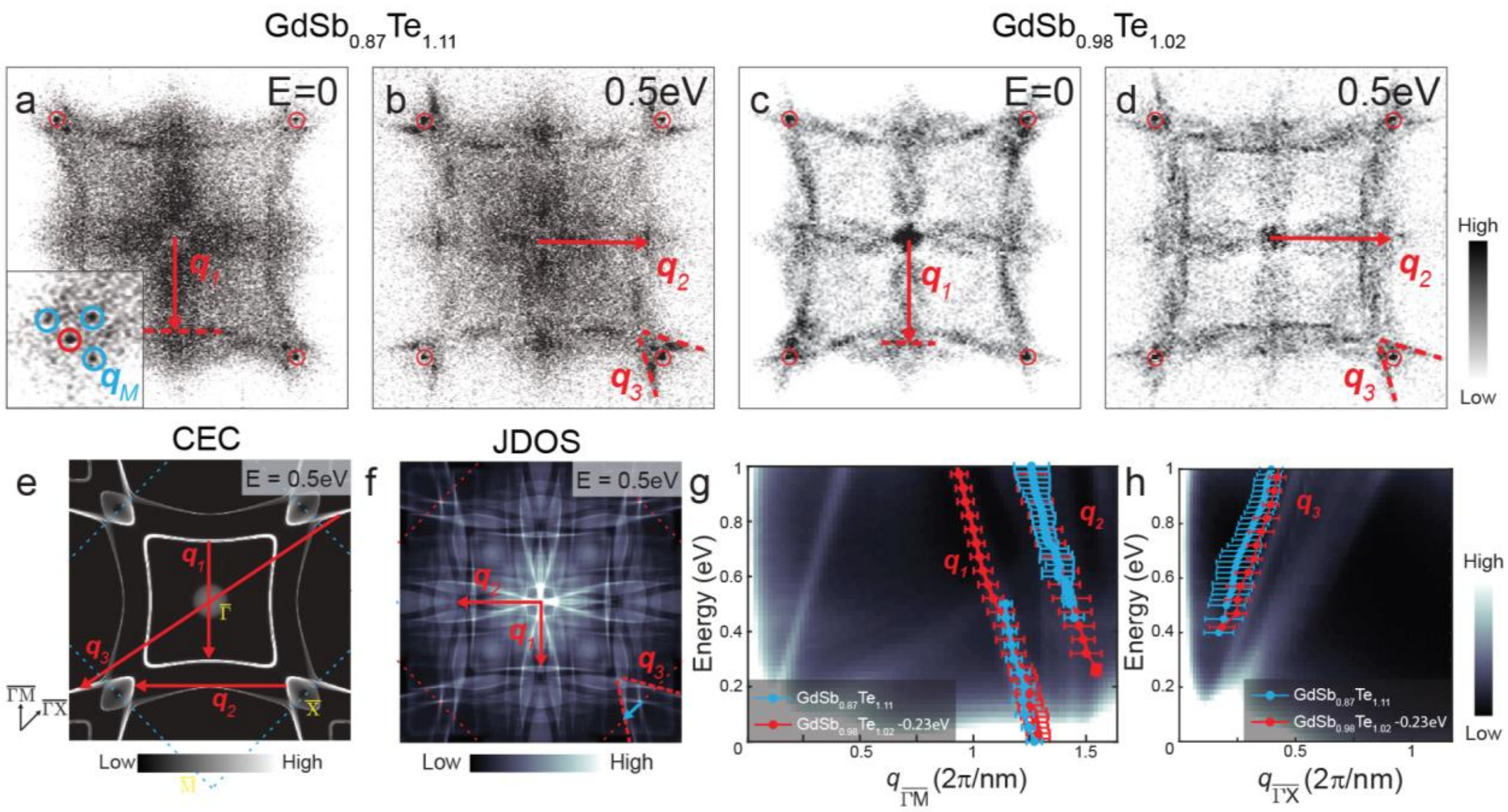

**Fig. 2. QPI imaging of $GdSb_{0.87}Te_{1.11}$ and $GdSb_{0.98}Te_{1.02}$ at T = 4.2 K. a - d** $L(\boldsymbol{q}$, E = 0 and 0.5 eV) maps of $GdSb_{0.87}Te_{1.11}$ and $GdSb_{0.98}Te_{1.02}$ obtained by Fourier transform of corresponding $L(\boldsymbol{r}$, E) maps in Fig S3. The red arrows indicate dispersive $\boldsymbol{q}$-vectors, $\boldsymbol{q_1}$, $\boldsymbol{q_2}$, and $\boldsymbol{q_3}$ and the red circles label the Bragg peaks. The inset in (**a**) shows the zoom-in image of the supermodulation, $\boldsymbol{q_M}$ (cyan circles) around Bragg peaks, which is visible around E > -0.3 eV. **e** Calculated constant energy contours (CEC) at E = 0.5 eV and the scattering process of $\boldsymbol{q_1}$, $\boldsymbol{q_2}$ and $\boldsymbol{q_3}$. The dashed line is the boundary of the first Brillouin zone. **f** Calculated joint density of states at E = 0.5 eV (details in Fig. S4). **g**, **h** Comparison of the $\boldsymbol{q}$-vector dispersion between the QPI data and the JDOS calculation (background) along $\overline{\Gamma M}$ direction and along the cyan arrow in (**f**), respectively. The cyan and the red dots represent the data from of $GdSb_{0.87}Te_{1.11}$ and $GdSb_{0.98}Te_{1.02}$. Energy in DFT calculated images, (**e**) ~ (**h**) are adjusted to fit the Fermi level of $GdSb_{0.87}Te_{1.11}$. Parameters for the $dI/dV(\boldsymbol{r}$, E) maps of $GdSb_{0.87}Te_{1.11}$ are 400 × 400 pixels, setpoint: V = -300 mV, I = 200 pA and lock-in modulation, $V_{mod}$ = 5 mV for E ≤ 500 meV and $V_{mod}$ = 20 mV for E > 500 meV and those for $GdSb_{0.98}Te_{1.02}$ are 300 × 300 pixels, setpoint: V = 500 mV, I = 200 pA for E ≤ 500 meV and V = 900 mV, I = 250 pA for E > 500 meV with $V_{mod}$ = 15 mV.

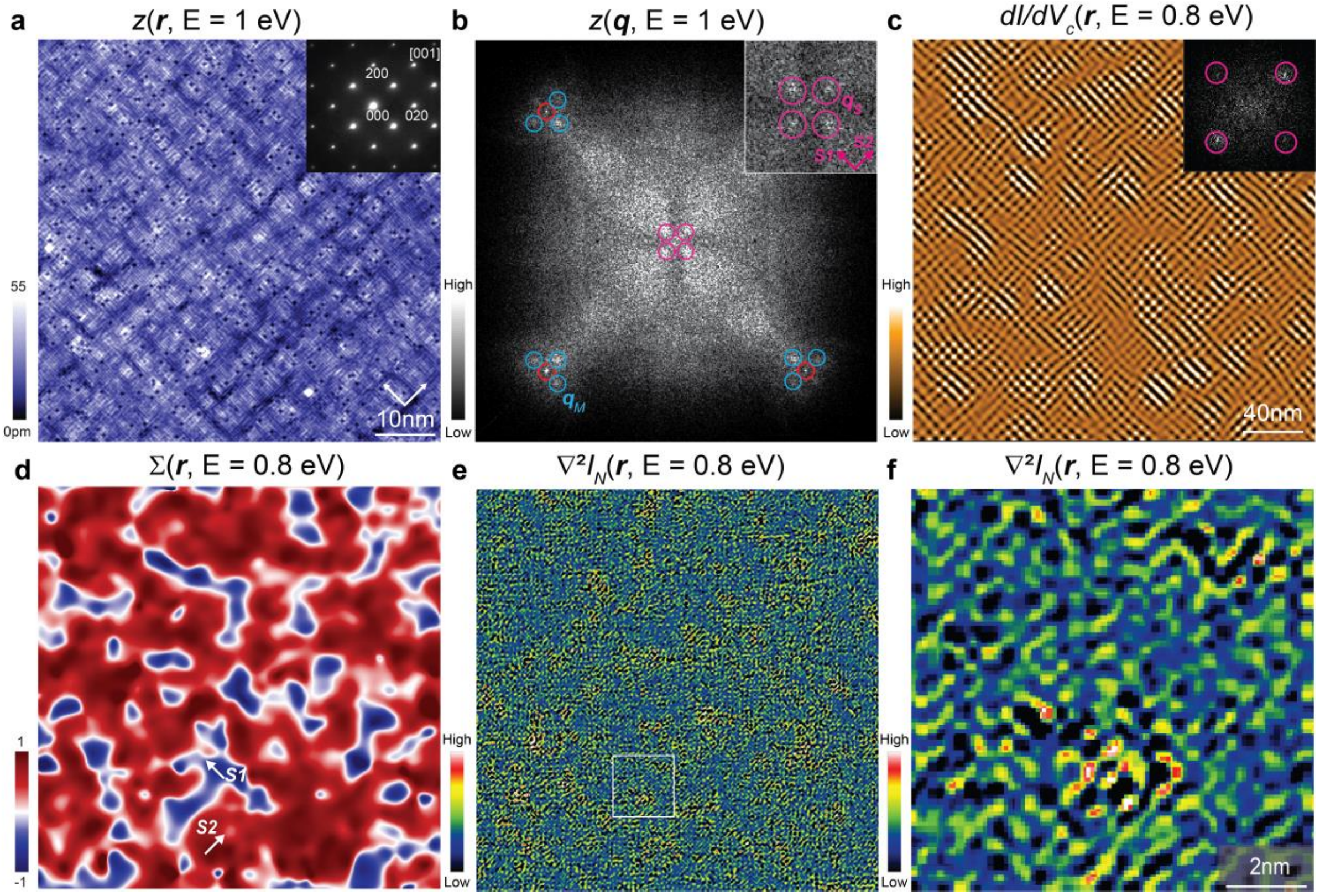

**Fig. 3. Electronic liquid crystal phases of $GdSb_{0.87}Te_{1.11}$ at T = 4.2 K. a** Topographic image of $GdSb_{0.87}Te_{1.11}$, $z(\boldsymbol{r}$, E = 1 eV) acquired in the same FOV as in Fig. 1**e** (I = 20 pA). The inset shows the electron diffraction pattern measured in the [001] zone of tetragonal $GdSb_{0.87}Te_{1.11}$ at T = 100 K. No other lattice modulation or distortion is observed. **b** Fourier transformation of the topograph in **a**. The red, magenta and cyan circles indicate Bragg peaks, smectic peaks ($\boldsymbol{q_S} = 2\pi/12.3\mathbf{a}$) and supermodulation peaks ($\boldsymbol{q_M} = 2\pi/\mathbf{a} \pm \boldsymbol{q_S}$), respectively. The inset shows the enlarged area at the center. We assign *S1* and *S2* for two Te-Te directions as denoted by two magenta arrows. **c** Electronic checkerboard pattern, $dI/dV_c(\boldsymbol{r}$, E = 0.8 eV), obtained from inversed Fourier transform of $\boldsymbol{q_s}$ in unprocessed $dI/dV(\boldsymbol{q}$, E = 0.8 eV) (the inset). The image was taken in a larger FOV of 240 × 240 $nm^2$ (setpoint: V = -0.3 V, I = 0.2 nA, $V_{mod}$ = 20 mV, details in Fig. S5 ~ S7). Broken $C_4$-rotational symmetry of $\boldsymbol{q_s}$ is evident from $dI/dV(\boldsymbol{q}$, E = 0.8 eV) (the inset). Thus, the charge modulation breaks both translational and rotational symmetry, representing a smectic phase. **d** Local Ising-like order parameter map, $\Sigma(\boldsymbol{r}$, E = 0.8 eV). Red (blue) areas indicate the charge modulation favors toward $S_2$($S_1$) direction as denoted by the white arrows. The limit of ±1 indicates pure

stripes. **e** Laplacian enhanced electronic unidirectional nanostructure, $\nabla^2 I_N$ (***r***, E = 0.8 eV) acquired by filtering all periodic signals in *I*(***r***, E = 0.8 eV). The FOV is identical as in Fig. 3**a** (setpoint: V = -0.3 V, and I = 0.2 nA). **f** The enlarged image of electronic unidirectional nanostructure taken in the area marked with the white box in **e**. The regions with higher intensity show clear unidirectionality.

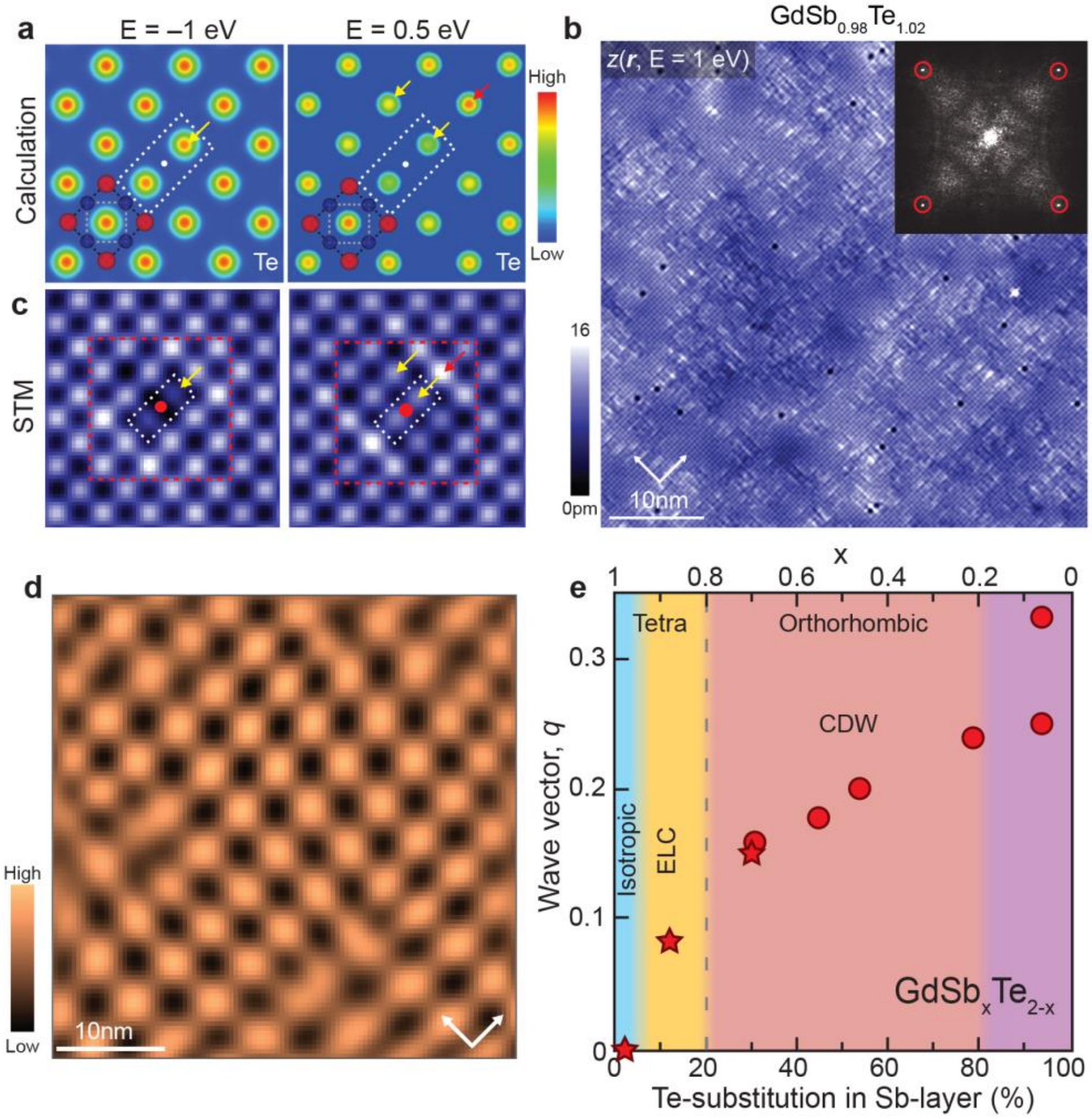

**Fig. 4. $C_2$ impurity state and the origin of ELC phase. a** Calculated charge density distribution of the Te-surface at E = -1 eV and E = 0.5 eV. The white dashed rectangle shows two nearest Te-atoms with lower $C_2$-symmetric charge density when a Te atom replaces a Sb atom ($Te_{Sb}$) in the Sb-layer below at the location marked with a white dot. Yellow and red arrows mark Te atoms with lower and higher charge density, respectively. **b** Topographic image of $GdSb_{0.98}Te_{1.02}$ acquired at T = 4.2 K (V = 1 V, I = 20 pA). Local unidirectional patterns are visibly aligned along Te-Te direction in proximity with $C_2$ impurity sites (details in Fig. S15). Fourier analysis shows only Bragg peaks (red circles) and residual QPI without $\boldsymbol{q_S}$ and $\boldsymbol{q_M}$ (inset). **c** Averaged topographic images taken with E = -1 eV and 0.5 eV, I = 100 pA at a $Te_{Sb}$-site (details in Fig. S13 and S15). The red box indicates the same location in (**a**). **d** Simulated spatial density distribution of Ising-like order parameters. **e** Proposed phase diagram of $GdSb_xTe_{2-x}$, based on our data (red stars) and Ref.[11,12] (red circles). The observed ELC phase is likely a precursor of CDW formation.

# Supplementary Information

## Direct Visualization of a Disorder Driven Electronic Smectic Phase in Nonsymmorphic Square-Net Semimetal GdSbTe

Balaji Venkatesan,[1,2,3]§ Syu-You Guan[3]§, Jen-Te Chang[3], Shiang-Bin Chiu[1], Po-Yuan Yang[4], Chih-Chuan Su[3], Tay-Rong Chang[4,13,14], Kalaivanan Raju[3], Raman Sankar[3], Somboon Fongchaiya[5,6,12], Ming-Wen Chu[5], Chia-Seng Chang[1,2,3], Guoqing Chang[7], Hsin Lin[3], Adrian Del Maestro[8,9,10], Ying-Jer Kao[1,11]*, Tien-Ming Chuang[3]*

[1] Department of Physics, National Taiwan University, Taipei 10617, Taiwan
[2] Nano Science and Technology Program, Taiwan International Graduate Program, Academia Sinica and National Taiwan University, Taipei 11529, Taiwan
[3] Institute of Physics, Academia Sinica, Taipei 11529, Taiwan
[4] Department of Physics, National Cheng-Kung University, Tainan 70101, Taiwan
[5] Center for Condensed Matter Sciences, National Taiwan University, Taipei 10617, Taiwan
[6] Molecular Science and Technology Program, Taiwan International Graduate Program, Academia Sinica, Taipei 11529, Taiwan
[7] Division of Physics and Applied Physics, School of Physical and Mathematical Sciences, Nanyang Technological University, 21 Nanyang Link 637371, Singapore
[8] Department of Physics and Astronomy, University of Tennessee, Knoxville, TN 37996, USA
[9] Institute for Advanced Materials & Manufacturing, University of Tennessee, Knoxville, TN 37996, USA
[10] Min H. Kao Department of Electrical Engineering and Computer Science, University of Tennessee, Knoxville, TN 37996, USA
[11] Center for Quantum Science and Technology, National Taiwan University, Taipei 10607, Taiwan
[12] International Graduate Program of Molecular Science and Technology (NTU-MST), National Taiwan University, Taipei 10617, Taiwan
[13] Center for Quantum Frontiers of Research and Technology (QFort), Tainan 70101, Taiwan
[14] Physics Division, National Center for Theoretical Sciences, Taipei 10617, Taiwan

§ These authors contributed equally to this work.
*Corresponding author
Email: yjkao@phys.ntu.edu.tw; chuangtm@gate.sinica.edu.tw

**This PDF file includes:**

Figs. S1 to S19
Supplementary Note 1 and 2
Table S1

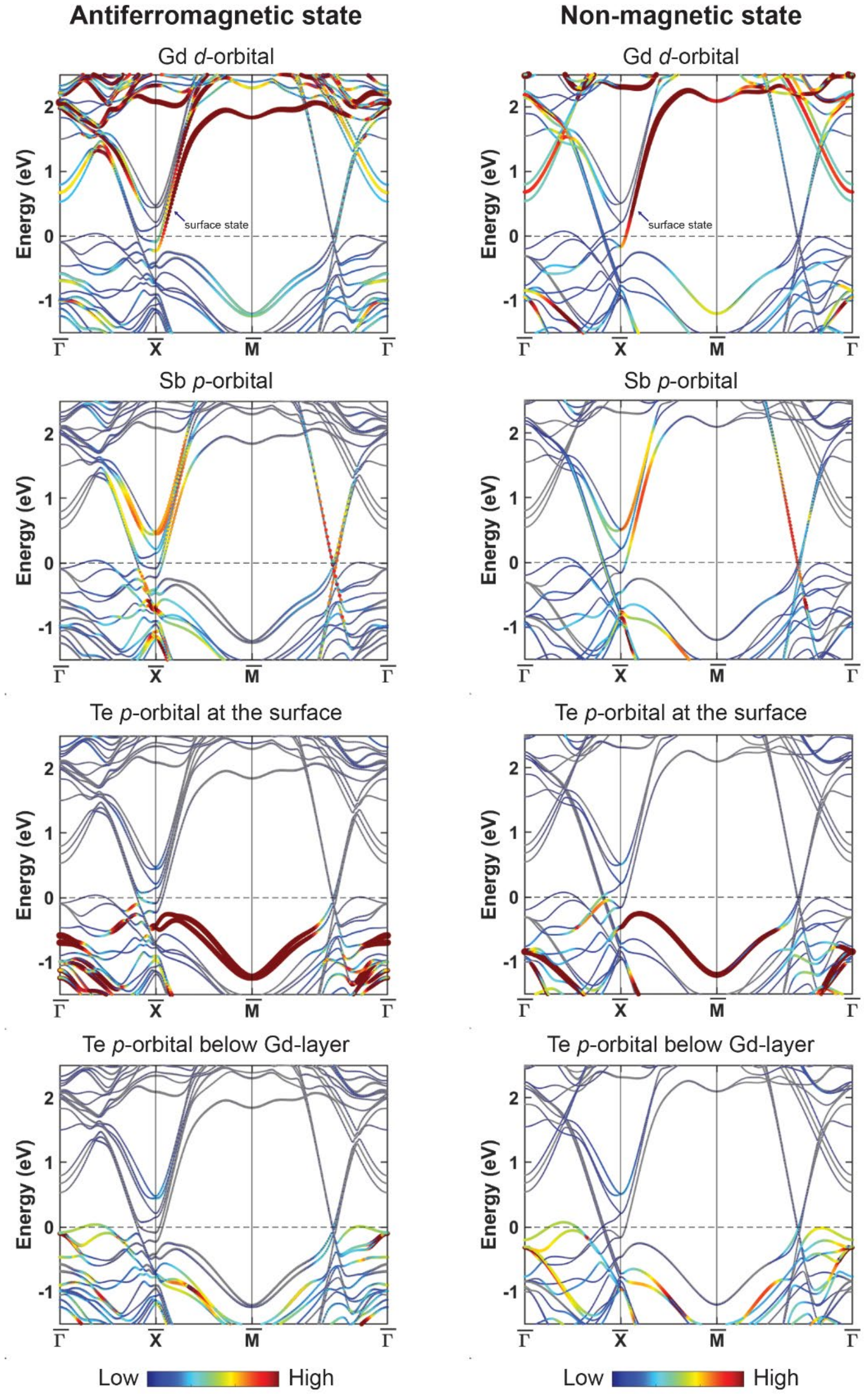

**Fig. S1.** Surface electronic structure of GdSbTe by slab calculation.

Surface electronic structure in the antiferromagnetic and non-magnetic state with orbital contributions from different constituted atoms in the unit cell.

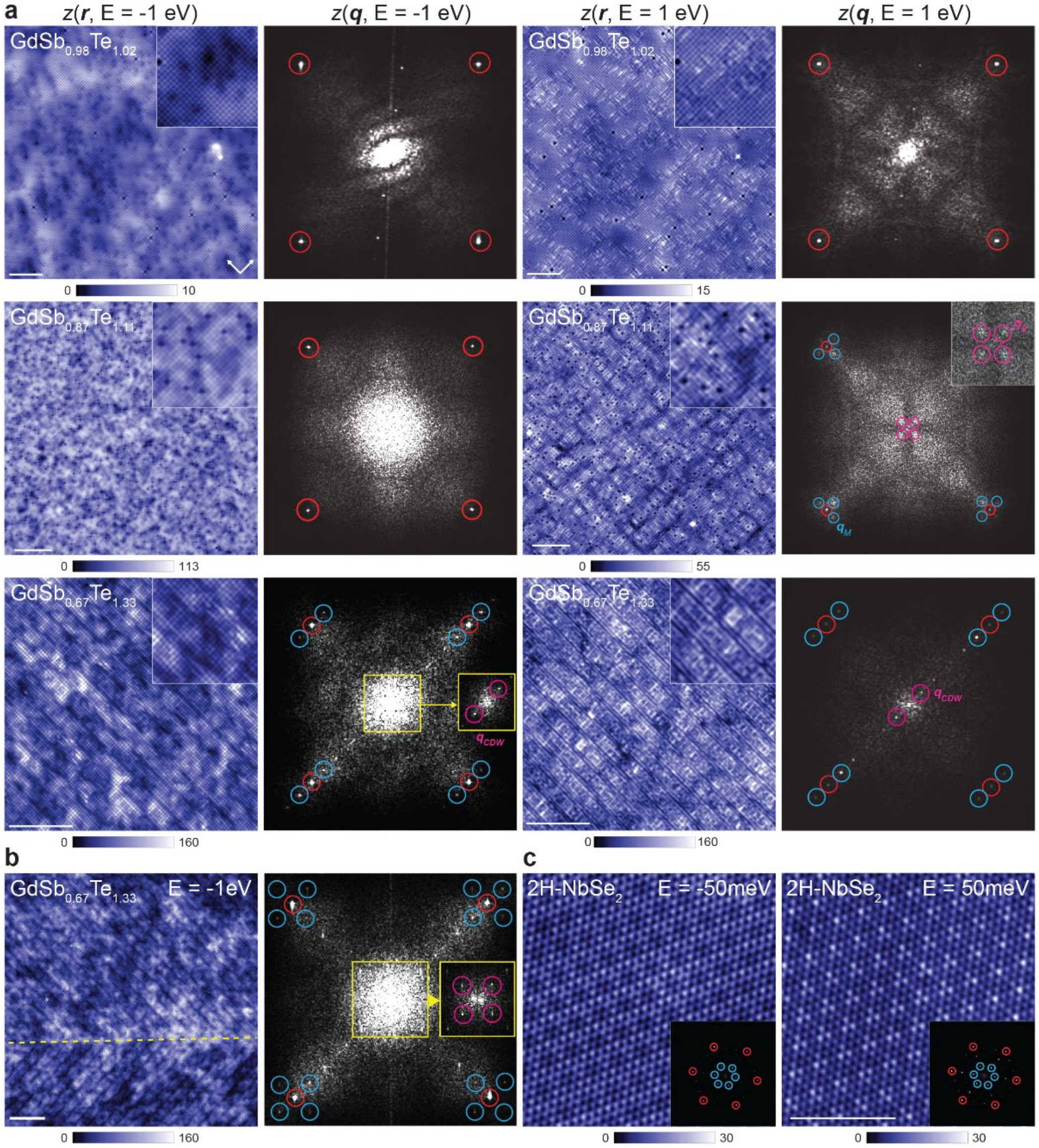

**Fig. S2.** Overview of STM topographic images of $GdSb_xTe_{2-x}$ (x = 0.98, 0.87 and 0.67). **a** topographic images, $z(\boldsymbol{r}, E)$ and their corresponding Fourier analysis taken at E = -1 eV (left) and E = 1 eV (right). For the same x, the FOV is identical. The white scale bar represents 10nm. The FOV is 10 × 10 nm$^2$ for all insets. Bragg peaks and supermodulation peaks ($\boldsymbol{q_M}$) are denoted by red and cyan. The smectic charge modulation ($\boldsymbol{q_S}$) and CDW ($\boldsymbol{q_{CDW}}$) peaks are labeled with magenta peaks. $GdSb_{0.67}Te_{1.33}$ exhibits a CDW wavelength of ~ 6.5**a**, which bares striking similarity with underdoped cuprates, is consistent with recent report of bond density wave in the CDW phase of CeSbTe in *Ref.* 31. **b** Topograph of $GdSb_{0.67}Te_{1.33}$ taken at a twin domain boundary (yellow dashed line) and its FFT analysis reveal the CDW modulation rotates 90° across the boundary. **c** Topographs of 2H-$NbSe_2$, where both Bragg (red) and CDW peaks (cyan) exist at E = ± 50 meV. We note that $\boldsymbol{q_{CDW}}$ show no bias dependence in $GdSb_{0.67}Te_{1.33}$ and 2H-$NbSe_2$. Similarly, STM topographs show the same atomic structure at different bias voltages in the case of surface reconstruction such as Si(111)-7×7. In contrast, $\boldsymbol{q_S}$ is only visible in $z(\boldsymbol{r}, E = 1 \text{ eV})$ in $GdSb_{0.87}Te_{1.11}$ whereas $\boldsymbol{q_S}$ is completely absent in $GdSb_{0.98}Te_{1.02}$. All images here are taken at T = 4.2 K.

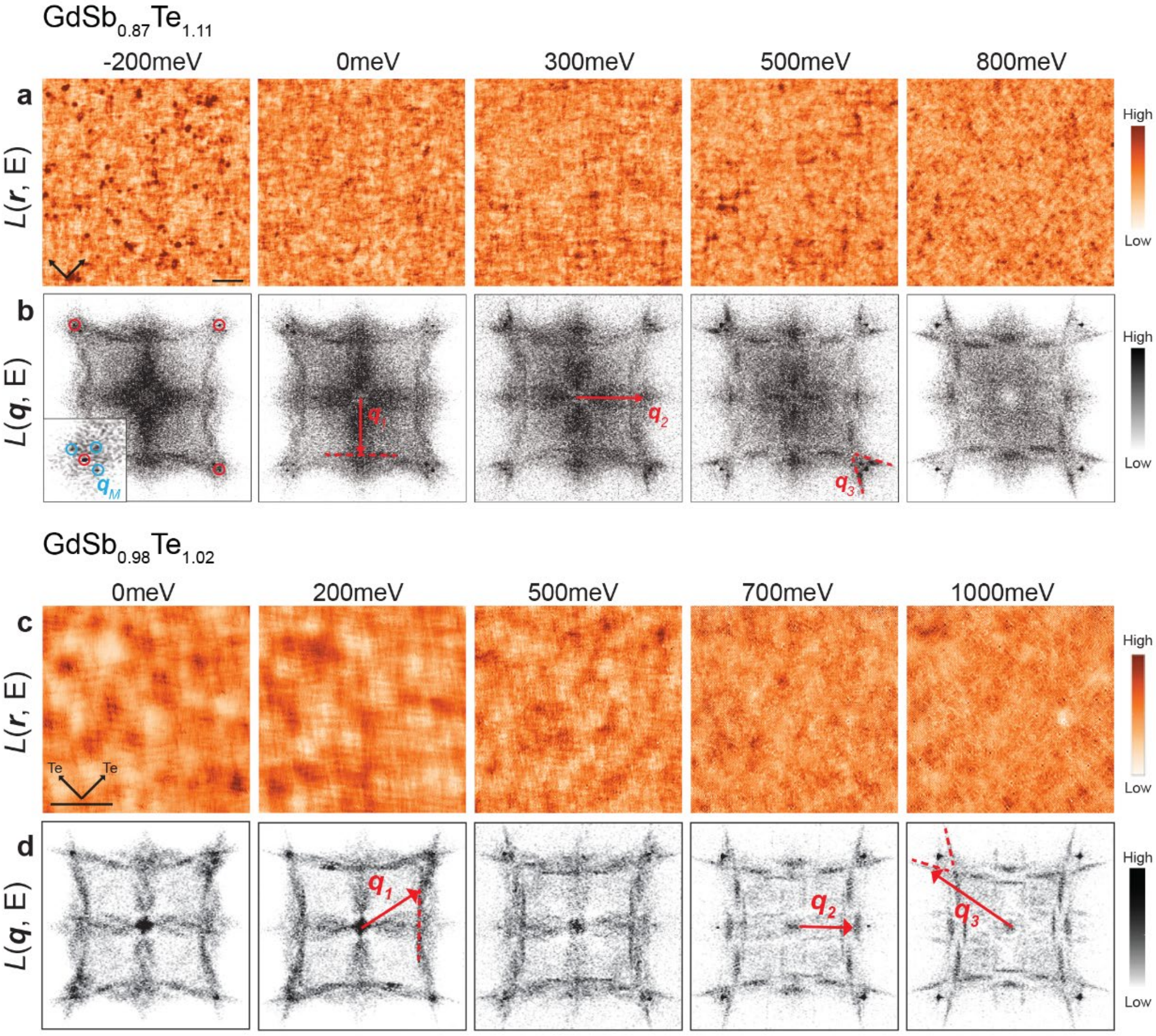

**Fig. S3.** Normalized differential conductance maps and QPI images of tetragonal $GdSb_xTe_{2-x}$ at T = 4.2 K.

**a** Normalized differential conductance maps, which is defined as $L(\boldsymbol{r}, \mathrm{E}) \equiv dI/dV(\boldsymbol{r}, \mathrm{E})/(I(\boldsymbol{r}, \mathrm{E})/V)$, for $GdSb_{0.87}Te_{1.11}$ (400 × 400 pixels, setpoint: V = -300 mV, I = 200 pA and lock-in modulation, $V_{mod}$ = 5 mV for E ≤ 500 meV and $V_{mod}$ = 20 mV for E > 500 meV). Such a normalization is used to minimize the setpoint effect, which can create artifacts in *dI/dV* measurements. Fig. S19 shows the detailed comparison between the differential conductance maps with and without normalization, in which no extra signal is observed. **b** Discrete Fourier transform of $L(\boldsymbol{r}, \mathrm{E})$ maps, defined as $L(\boldsymbol{q}, \mathrm{E})$, at the corresponding energies. **c** Normalized differential conductance maps, $L(\boldsymbol{r}, \mathrm{E})$ for $GdSb_{0.98}Te_{1.02}$ (300 × 300 pixels, setpoint: V = 500 mV, I = 200 pA for E ≤ 500 meV and V = 900 mV, I = 250 pA for E > 500 meV with $V_{mod}$ = 15 mV). **d** Discrete Fourier transform of $L(\boldsymbol{r}, \mathrm{E})$ maps, $L(\boldsymbol{q}, \mathrm{E})$ at the corresponding energies. The scale bars in (**a**) and (**c**) represent 10 nm and the arrows indicate the Te-Te direction. Zero frequency noises are suppressed for clarity in (**b**) and (**d**). All QPI images are not symmetrized.

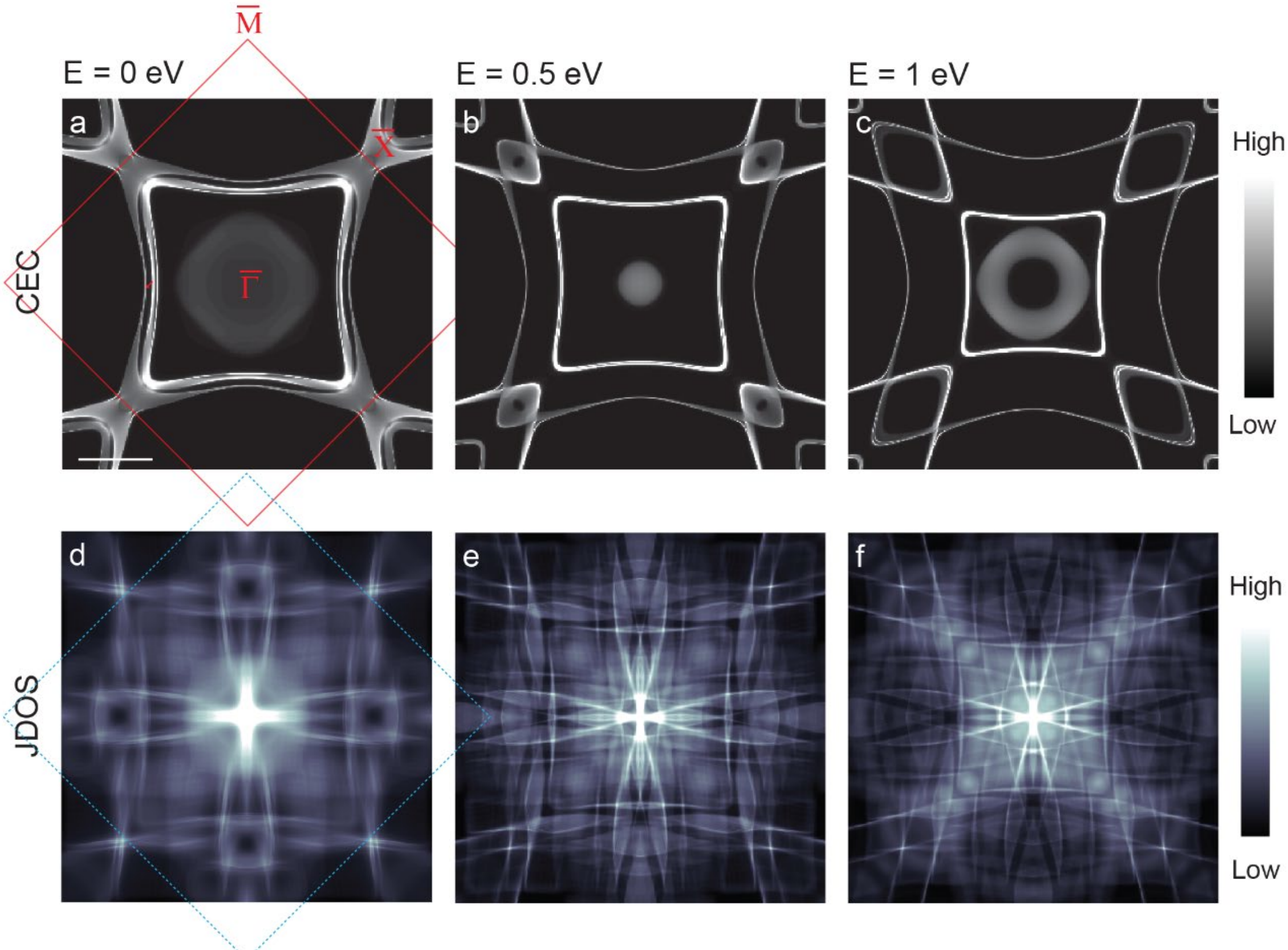

**Fig. S4.** Simulated QPI patterns by JDOS calculations.

**a** – **c** The constant energy contours (CECs) of surface band structure by DFT slab calculation (Fig. S1) at E = 0, 0.5 eV and 1 eV, respectively. The scale bar represents π/nm. The Fermi level here is adjusted to fit the experimental value from QPI fitting of $\boldsymbol{q_1}$ in $GdSb_{0.87}Te_{1.11}$. **d** – **f** The simulated QPI images at the corresponding energies by JDOS calculations in (**a**) - (**c**). The red and cyan box indicate the first Brillouin zone and $\pm 2\pi/\boldsymbol{a}$, respectively. A wavevector of ~ 2π/14.8**a** connecting the inner and the outer diamond band is indicated by the red arrow in (A). While the nesting vector between the two bands is associated with CDW. The magnitude which does not agree with the smectic charge modulation of ~12.3**a** in this work.

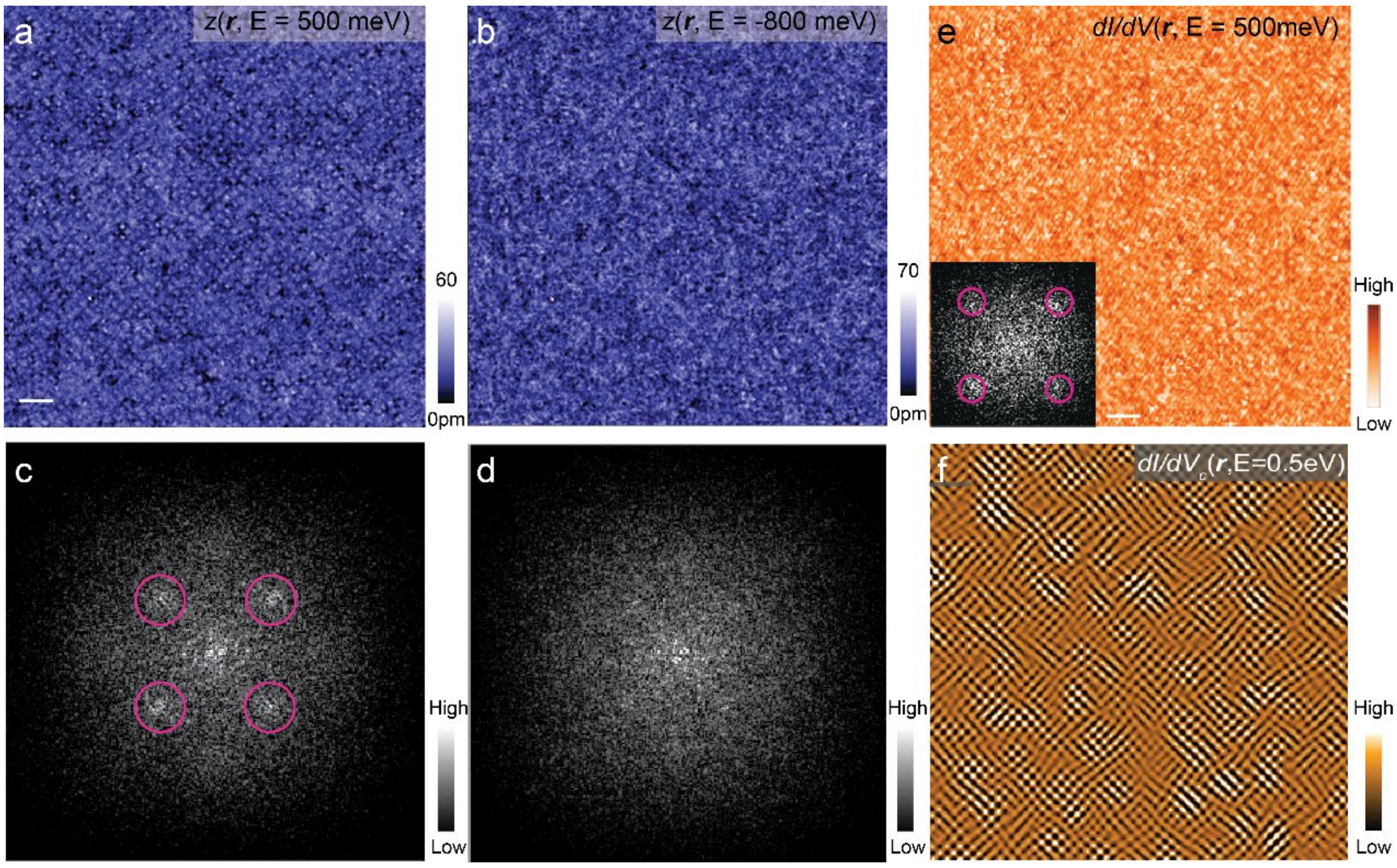

**Fig. S5.** Translational symmetry breaking in a FOV of 240 × 240 nm$^2$ at T = 4.2 K. The analysis of $\boldsymbol{q_S}$ can be hindered by the noise at $\boldsymbol{q}$ = 0. Images taken in a large FOV can zoom in around $\boldsymbol{q}$ = 0 in $\boldsymbol{q}$-space and mitigate this issue as shown here. STM topographic images taken at a bias voltage of **(a)** 500 mV and **(b)** -800 mV, respectively (I = 3 pA). The scale bar represents 20 nm. Their corresponding FFT images are shown in **(c)** and **(d)** in which $\boldsymbol{q_S}$ is marked with magenta circles. **(e)** Conductance map, *dI/dV*(***r***, E = 0.5 eV) taken at E = 0.5 eV (setpoint: V = -500 mV, I =200 pA, $V_{mod}$ = 50 mV) and its Fourier analysis (inset). **(f)** Real space charge modulation of $\boldsymbol{q_S}$, $dI/dV_C$ (***r***, E = 0.5 eV). We first applied a mask of a 2D boxcar function (amplitude = 1 and a diameter of 0.15π/nm) to each $\boldsymbol{q_S}$ in *dI/dV* (***q***, E) (indicated by the magenta circles in the inset of Fig. **S5e**), yielding a masked image with four isolated $\boldsymbol{q_S}$. We then obtained $dI/dV_C$ (***r***, E) by taking inverse Fourier transform of the masked *dI/dV*(***q***, E) image.

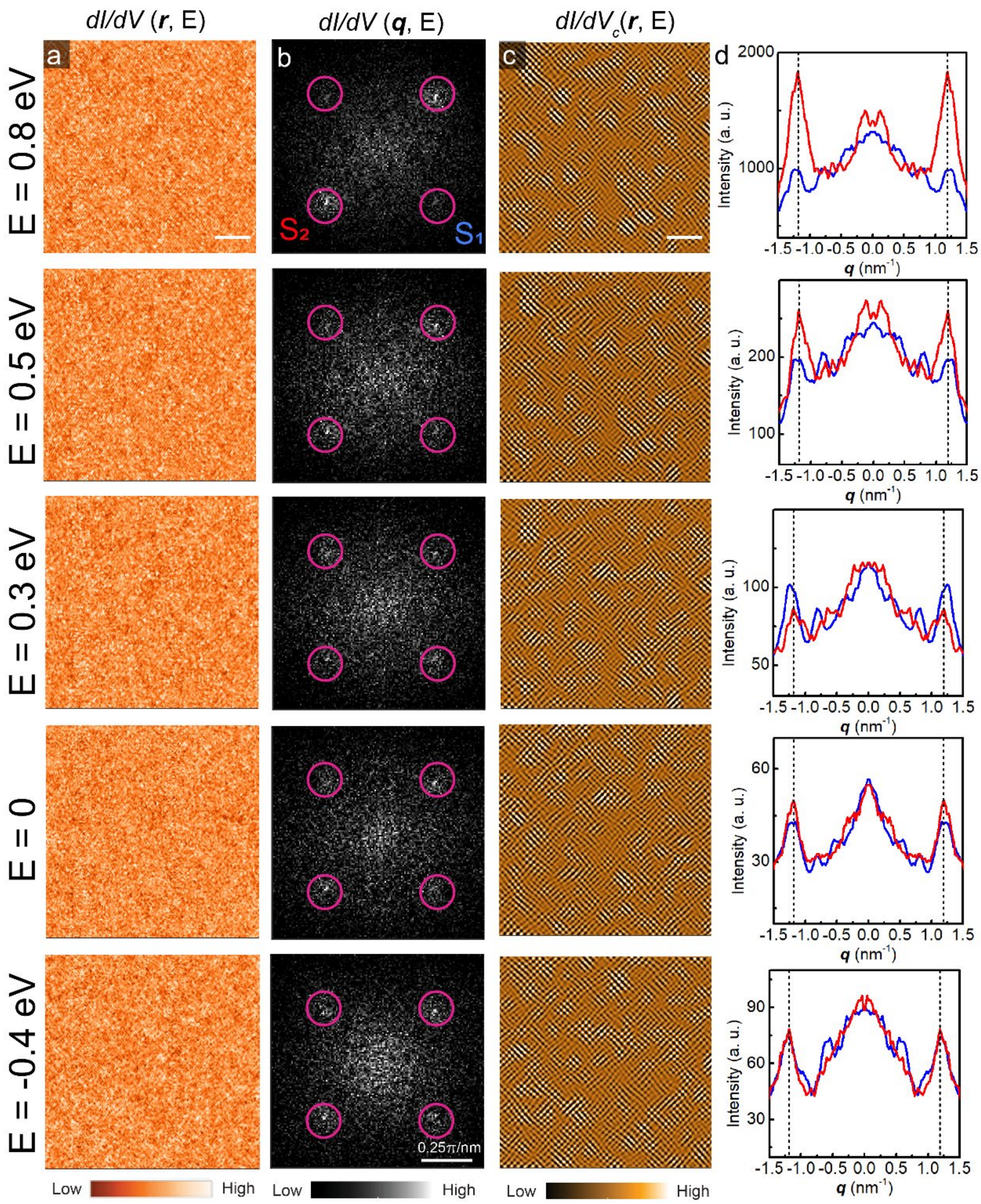

**Fig. S6**. The evolution of $\boldsymbol{q_S}$ as a function of energy at T = 4.2 K.

**a** Differential tunneling conductance maps taken in the same FOV as in Fig. S5. **b** Fourier analysis of conductance maps in (**a**) reveals the smectic charge modulation, $\boldsymbol{q_S}$ at corresponding energies. $\boldsymbol{q_S}$ breaks rotational symmetry and becomes smectic at E ≥ 0.5 eV while it exhibits nearly $C_4$-symmetry at lower energy. **c** The corresponding spatial distribution of $\boldsymbol{q_S}$ in the same FOV. We used the same procedures as in Fig. **S5f** to obtain these $dI/dV_C$ (***r***, E) images. Magenta circles in **b** show the size of boxcar function for masking. The scale bar in (**a**) and (**c**) represents 40 nm (setpoint: V = -500 mV, I = 10 pA and $V_{mod}$ = 50 mV). **d** Linecuts of the intensity of $\boldsymbol{q_S}$ obtained from **b** show the broken $C_4$-rotational symmetry in $\boldsymbol{q_S}$ becomes more evident at higher energies, which is also clear from $z(r$, E = ±1 eV).

**Supplementary Note 1.** Spatial distribution of smectic charge modulation and correlation length analysis

We define a generic density following the phenomenological methodology by Del Maestro *et al.* in *Ref. 21*:

$$\delta\rho = Re\left[\Phi_x e^{iK_x\cdot r}\right] + Re\left[\Phi_y e^{iK_y\cdot r}\right] \text{ …. (1)}$$

where $K_x$ and $K_y$ are the dominant wavevectors of CDWs, which are chosen based on an analysis of STM data. $\Phi_x$ and $\Phi_y$ are complex order parameters, which describe the spatial modulation of the CDW. To find the tendency to form stripes towards *x* or *y*, we adapt the local order parameter (*Ref. 21, 22*):

$$\Sigma(r) = \frac{\left|\Phi_y(\boldsymbol{r})\right|^2 - |\Phi_x(\boldsymbol{r})|^2}{|\Phi_x(\boldsymbol{r})|^2 + \left|\Phi_y(\boldsymbol{r})\right|^2} \text{ …… (2)}$$

The value of $\Sigma(\boldsymbol{r})$ ranges between +1 and -1, where a positive (negative) $\Sigma(r)$ indicates that the intensity of $\boldsymbol{q}_{s2}$ is larger (smaller) than that of $\boldsymbol{q}_{s1}$. When $\Sigma(\boldsymbol{r})$ = +1 or -1, it indicates a pure stripe along *y* or *x*, respectively. When $\Sigma(\boldsymbol{r}) = 0$, it indicates a pure checkerboard order.

For Fig. 3**d**, we first decompose *dI/dV*($\boldsymbol{q}$, E) along two Te-Te directions (denoted as $S_1$ and $S_2$), using inversed Fourier transform to obtain $dI/dV_{S1}$($\boldsymbol{r}$, E) and $dI/dV_{S2}$($\boldsymbol{r}$, E) (Fig. S7). Then, we follow the method outlined in *Ref. 36* to extract $\Phi_x(\boldsymbol{r})$ and $\Phi_y(\boldsymbol{r})$. Specifically, *dI/dV*($\boldsymbol{r}$, E) is multiplied by the real-space modulation with a wavelength corresponding to $\boldsymbol{q}_{s1}$. Then, the resulting product is Fourier-transformed, low-pass filtered (using a Gaussian window of width = 0.22 π/nm) at $\boldsymbol{q} = 0$ , and then inverse Fourier-transformed back to real space, yielding $|\Phi_x(\boldsymbol{r})|^2$. The same procedure is applied along the other direction to obtain $|\Phi_y(\boldsymbol{r})|^2$.

To examine the effectiveness of $\Sigma(\boldsymbol{r})$, we apply a mask with $|\Sigma(r)| \leq 0.1$ to the $dI/dV_C$ data. The Fourier transform of this masked data reveals a pattern close to $C_4$ symmetry (Fig. S8**a**). On the other hand, when the mask is chosen with $\Sigma(r) > 0.1$ and $\Sigma(r) < -0.1$, the Fourier transformed images exhibit $C_2$ symmetry along $\boldsymbol{q}_{S2}$ and $\boldsymbol{q}_{S1}$, respectively (Fig. S8**b** and Fig. S8**c**). This indicates that $\Sigma(\boldsymbol{r})$ successfully partitions *dI/dV* into distinct regions, depending on the tendency towards $\boldsymbol{S_1}$ or $\boldsymbol{S_2}$.

We further perform the correlation length analysis on our STM data by using the method outlined by Del Maestro *et al.* in *Ref. 21*. First, we determine the average data points $n_p$ within a single period of length 12.3$a$ (with $a$ = 4.3 Å) in either *x* or *y* direction. The wave vector $\boldsymbol{q_S} = 2\pi/12.3\boldsymbol{a}$ is then used as the CDW wave vector $K_x$ and $K_y$ in *x* and *y* direction, respectively. Subsequently, we define a square box □, which includes the data points within 12.3$a$ × 12.3$a$ nm². The box is centered as close as possible to the data point located at *r*. we calculated the local density-density correlation function:

$$S_{\square}\ (r, r' - r") = < \delta\rho(r')\delta\rho(r'') >_{r'-r'' \in \square_r} \text{…. (3)}$$

and then perform a local discrete Fourier transform of $S_{\square}\ (r, r' - r")$

$$S_{\square}\ (r,k) = \frac{1}{n_p^2} \sum_{r' \in \square_r} S_{\square}\ (r,r') e^{-ik \cdot r'} \text{............}(4)$$

Finally, we calculate an effective local Ising-like order parameter:

$$\widetilde{\Sigma} = \frac{S_{\square}\ (r,K_x) + S_{\square}\ (r,-K_x) - S_{\square}\ (r,K_y) - S_{\square}\ (r,-K_y)}{S_{\square}\ (r,K_x) + S_{\square}\ (r,-K_x) + S_{\square}\ (r,K_y) + S_{\square}\ (r,-K_y)} \text{.........}(5)$$

If the value of averaged Ising-like order parameter $|\bar{\bar{\Sigma}}|$ is smaller than 0.5, it indicates the checkerboard phase. Otherwise, it is in the stripe phase.

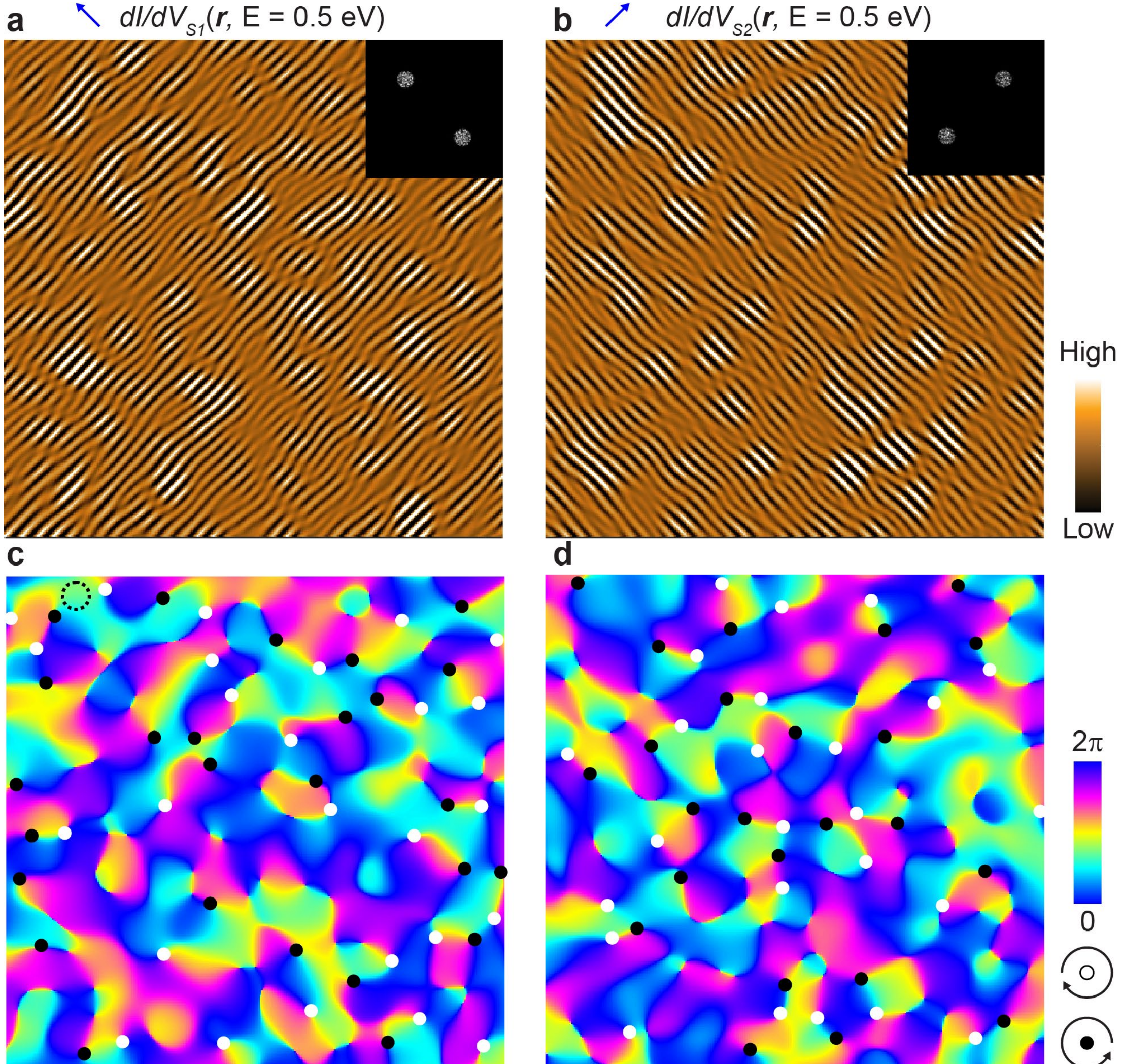

**Fig. S7.** Domains of stripes and topological defects.

The translational symmetry breaking charge modulation of $\boldsymbol{q_S}$ in Fig. S6 could be domains of stripes. **a** and **b** Spatial distribution of stripes along each Te-Te direction ($\boldsymbol{S_1}$ and $\boldsymbol{S_2}$ as indicated by the arrows), $dI/dV_{S1}(\boldsymbol{r})$ and $dI/dV_{S2}(\boldsymbol{r})$ can be visualized by inverse Fourier transform of $\boldsymbol{q_S}$ along $\boldsymbol{S_1}$ and $\boldsymbol{S_2}$ from $dI/dV(\boldsymbol{q})$ in Fig. S6b, respectively. The insets in **a** and **b** show the Fourier transforms of $\boldsymbol{q_{s1}}$ and $\boldsymbol{q_{s2}}$. The $dI/dV$ modulations along the direction of $\boldsymbol{q_{S1}}$ and $\boldsymbol{q_{S2}}$ correspond to $dI/dV_{S1}$ and $dI/dV_{S2}$, respectively. **c** and **d** Topological defects of each stripe domain are extracted from (**a**) and (**b**) by using the local 2D lock-in method (*Ref. 29*). The cutoff length is denoted by the dashed circle in (**c**) with a diameter of 12.5 nm.

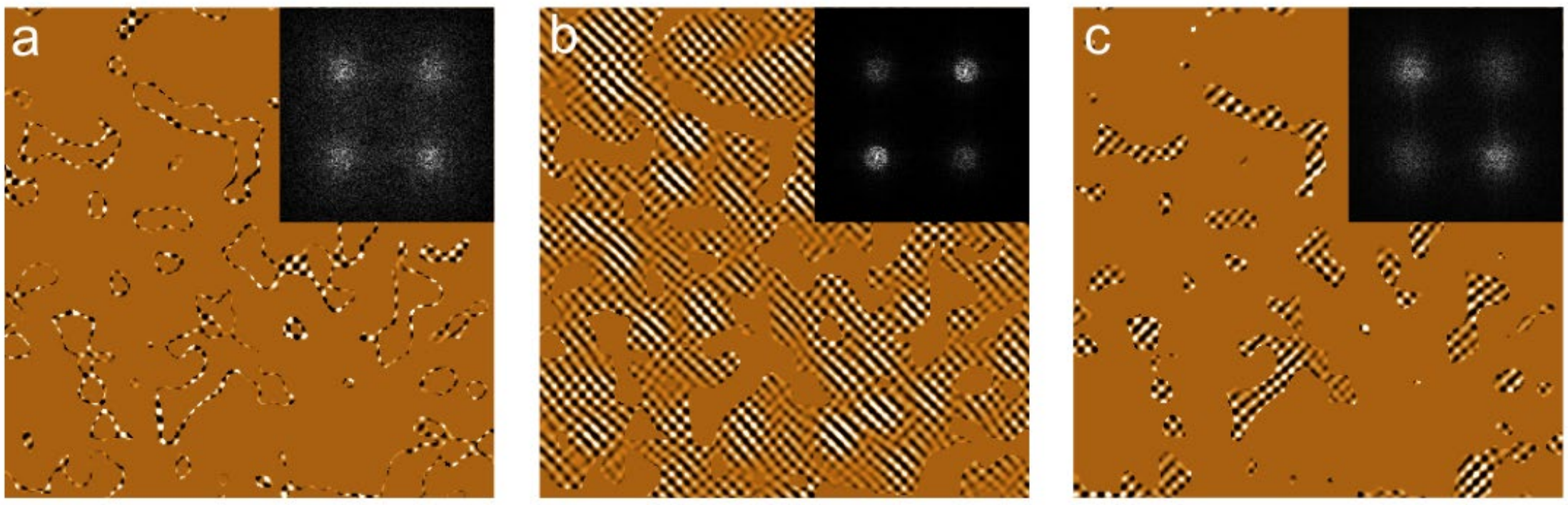

**Fig. S8.** Local stripeness in the smetic charge modulation.

**a**. The $dI/dV_C$ map under the mask of $|\Sigma(\boldsymbol{r})| \leq 0.1$. **b** and **c** The $dI/dV_C$ map under the mask of $\Sigma(\boldsymbol{r}) > 0.1$ and $\Sigma(\boldsymbol{r}) < -0.1$ shows regions with dominating $\boldsymbol{q_{S2}}$ and $\boldsymbol{q_{S1}}$, respectively. The insets show the Fourier transform of the corresponding masked maps, where $C_2$-symmetry is evident.

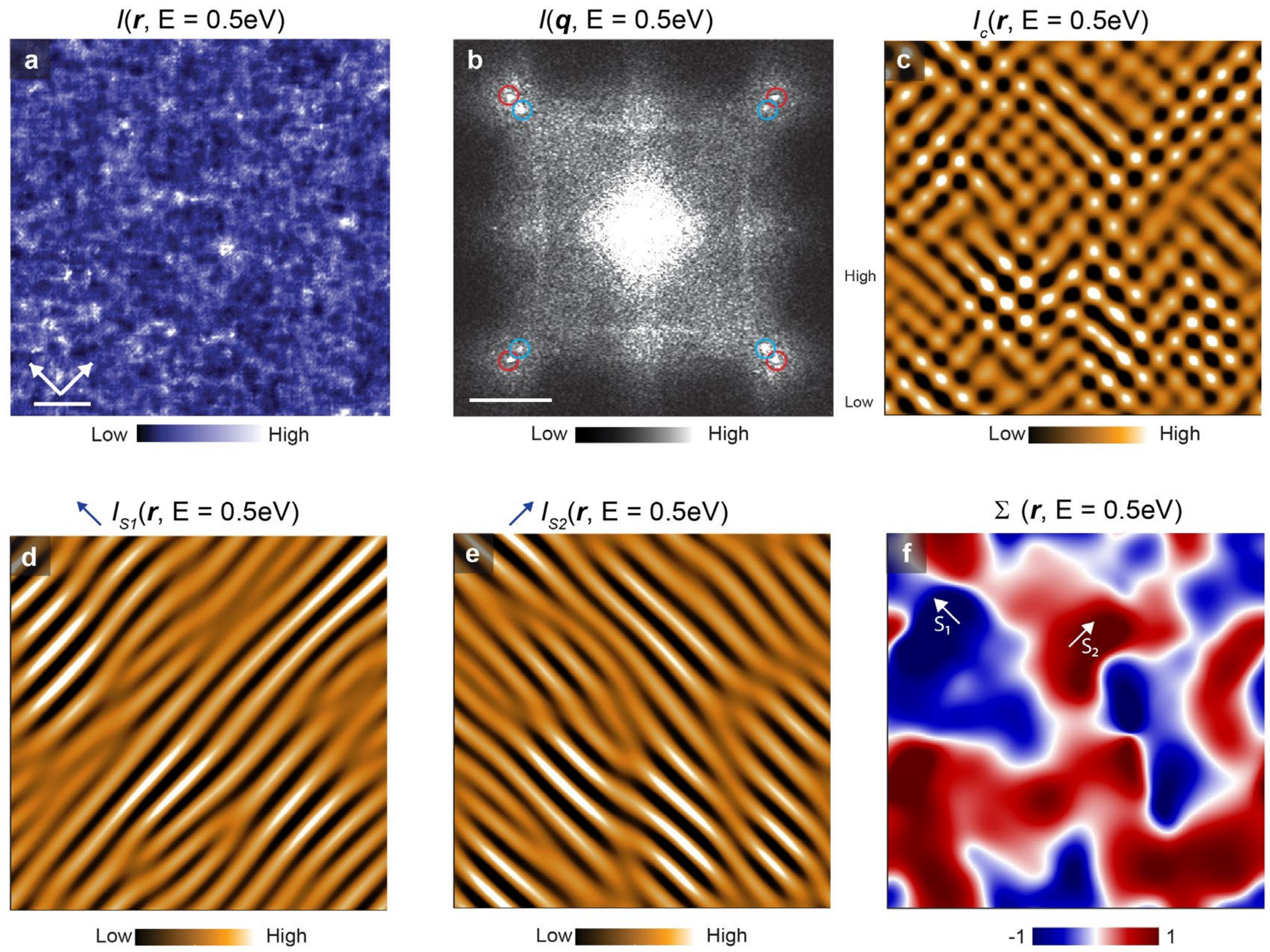

**Fig. S9.** Current map and local Ising-like order parameter map at T = 4.2 K.

**a** Current map, $I(\boldsymbol{r}$, E = 0.5 eV) acquired simultaneously with differential conductance maps in Fig. S3**a** and in the same FOV as in Fig. 1**e**, Fig. 3**a** and Fig. S3**a** (setpoint: V = -300 mV and I = 200 pA). The scale bar represents 10nm and the arrows indicate the Te-Te directions. **b** FFT image of $I(\boldsymbol{r}$, E = 0.5 eV) in (**a**). **c** Electronic checkerboard pattern, $I_c(\boldsymbol{r}$, E = 0.5eV), obtained from inversed Fourier transform of $\boldsymbol{q_S}$ in (**b**). **d** - **e** Spatial distribution of stripes along each Te-Te direction ($\boldsymbol{S_1}$ and $\boldsymbol{S_2}$ as denoted by the arrows). **f** Local Ising-like order parameter map, $\Sigma(\boldsymbol{r}$, E = 0.5eV), which is calculated from **d** and **e**, as defined in Supplementary Note 1.

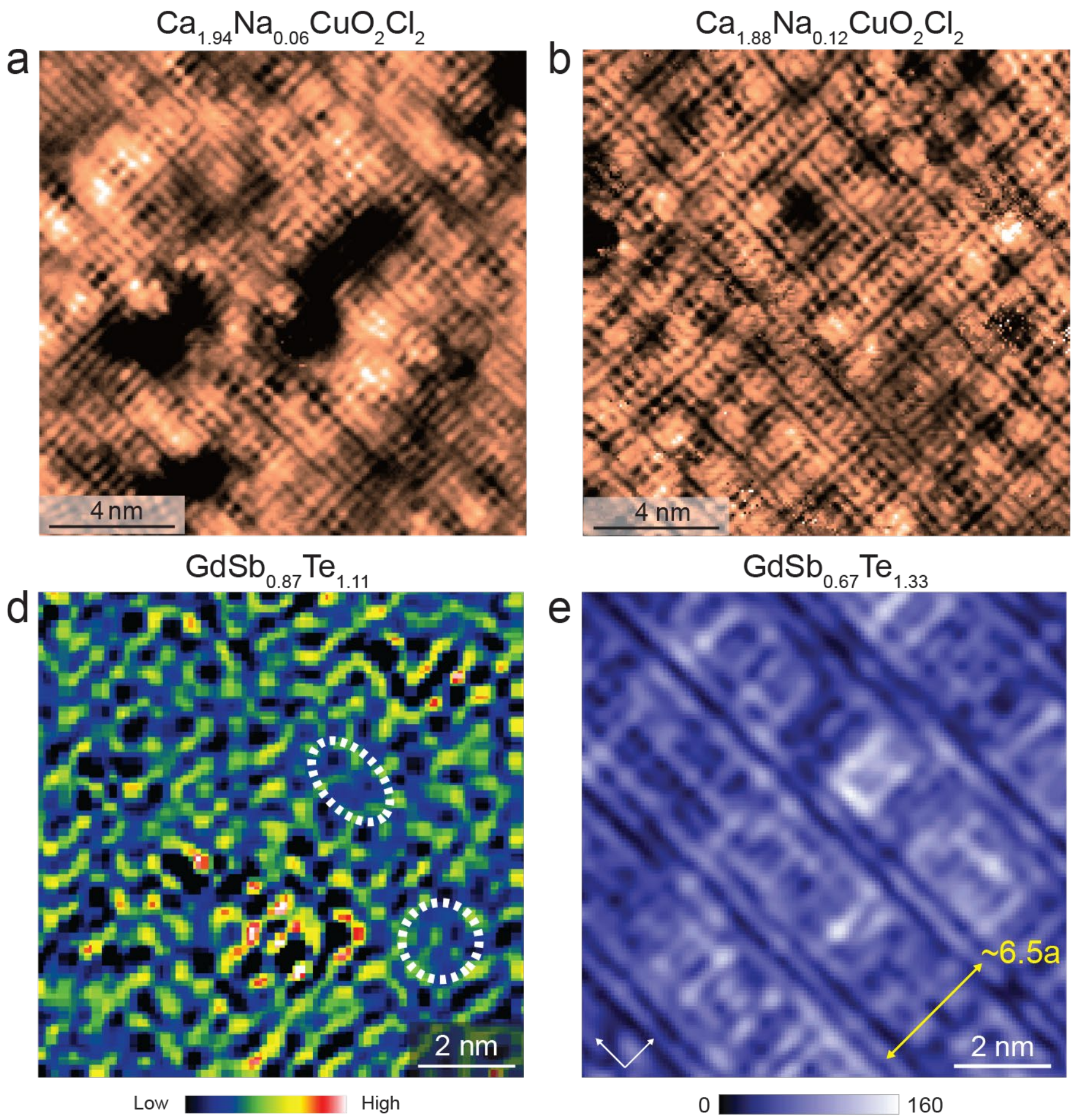

**Fig. S10.** Comparison between Na-doped $CaCuO_2Cl_2$ and $GdSb_xTe_{2-x}$.

**a** Current ratio map of 6% Na-doped and **b** 12% Na-doped $CaCuO_2Cl_2$ (adapted from the SI in *Ref.* 25 with permission from the corresponding author). $C_{2v}$-symmetric pseudogap and $C_{4v}$-symmetric insulating (dark) areas coexist at lower doping level as shown in **a** while $C_{2v}$-symmetric areas percolate at higher doping level. **c** Laplacian-enhanced Fourier-filtered current map, $\nabla^2 I_N(r, \mathrm{E} = 0.8\ \mathrm{eV})$ of $GdSb_{0.87}Te_{1.11}$ (as in Fig. 3**f**) shows unidirectional nanostructures (high intensity in current). Regions without broken symmetry (white circles) are also visible. **d** Topograph of $GdSb_{0.67}Te_{1.33}$ reveals the ladder-like structures with a width of approximately 6.5a (yellow arrow). The doping dependence is phenomenologically similar with underdoped cuprates.

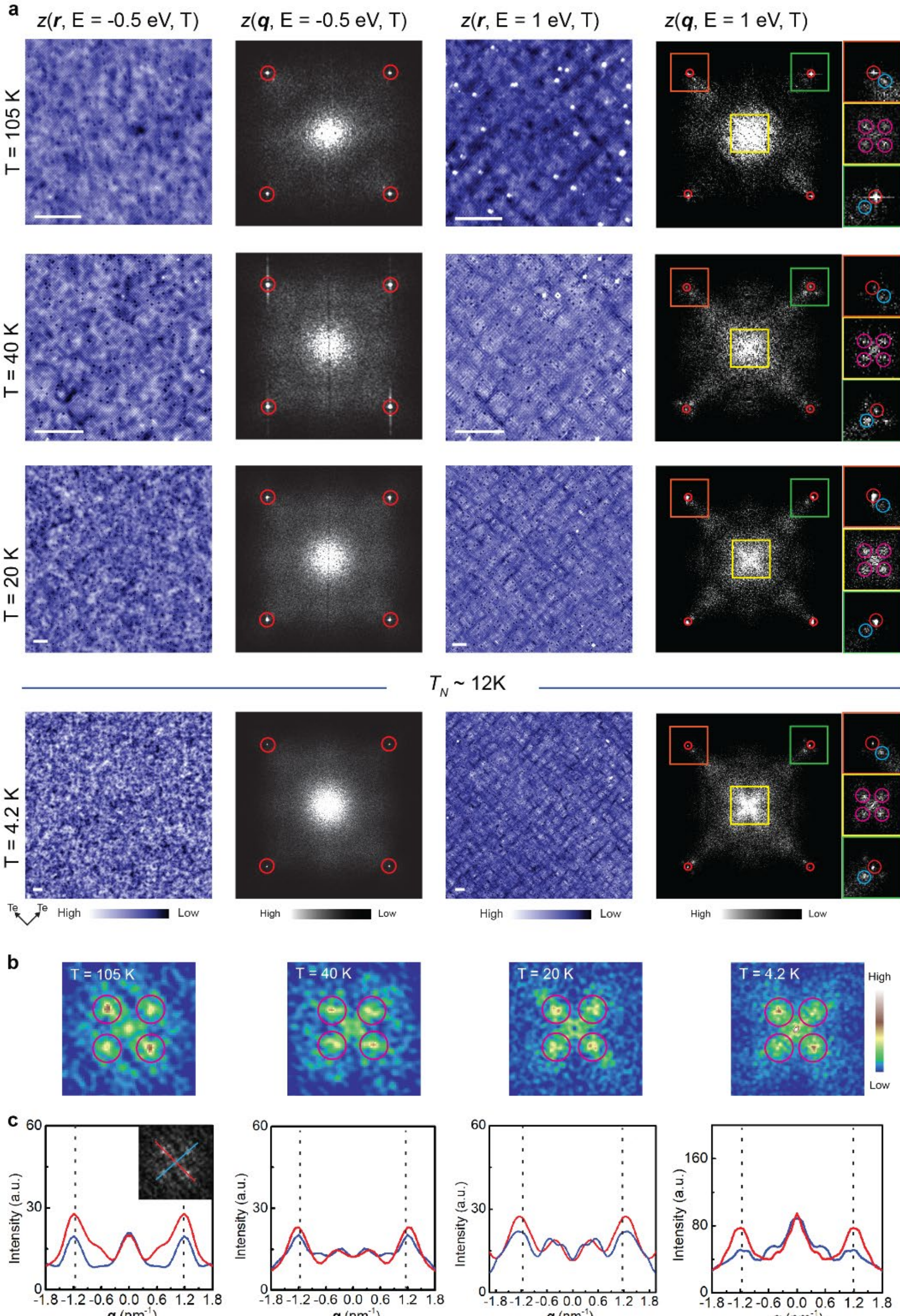

**Fig. S11.** Persistence of the ELC phase at elevated temperatures.

Topographic images, *z*(***r***, E, T) and their corresponding Fourier transformed images, *z*(***q***, E, T), taken at E = -0.5 eV and 1 eV in the same FOV from T = 4.2 K to T = 105 K. The scale bar represents 20 nm in all topographic images. Yellow, orange and green boxes show enlarged images from corresponding areas in *z*(***q***, E = 1eV, T) after adjusting the contrast. Bragg peaks, supermodulation peaks ($q_M$) and $q_S$, which are visible up to T = 105 K, are denoted by red, cyan and magenta peaks. Because the CDW transition in orthorhombic $GdSb_xTe_{2-x}$ occurs above room temperature, we speculate the transition temperature of the observed smectic phase here will also be very high, which is at the moment beyond the temperature range of our STM. **b** Enlarged images from yellow boxes in (**a**) after simple moving average of 1 pixel. **c.** Linecuts of the intensity of $q_S$ obtained from **a** show the four-fold rotational symmetry breaking in $q_S$.

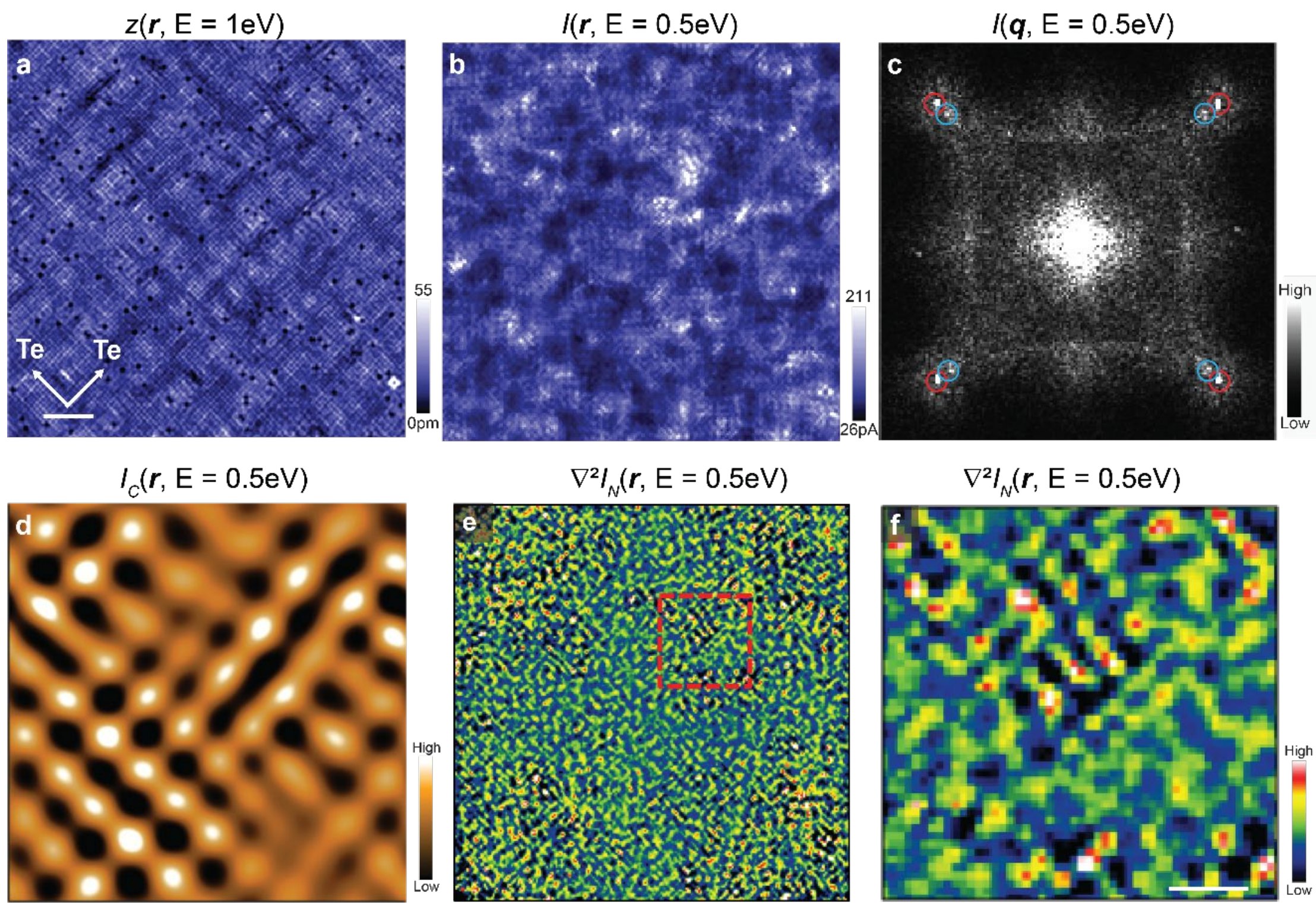

**Fig. S12.** The existence of the ELC phases in the non-magnetic state at T = 20 K.

**a** Topographic image taken at E = 1 eV (the scale bar represents 5 nm) and **b** Current map, $I(\boldsymbol{r}$, E = 0.5 eV) in the same FOV as (**a**) (setpoint : V = -300 mV, I = 200 pA). **c** The FFT of the image in (**b**) with the Bragg peak and the supermodulation marked in red and cyan circles. **d** Electronic checkerboard pattern obtained by inversed Fourier transform of $q_S$ in (**c**) along both Te-Te direction. **e** Laplacian enhanced electronic unidirectional nanostructure acquired by Fourier filtering all periodic signals in $I(\boldsymbol{r}$, E = 0.5eV). **f** Enlarged electronic unidirectional nanostructure from the area marked with the red box in (**e**). The scale bar indicates 2nm.

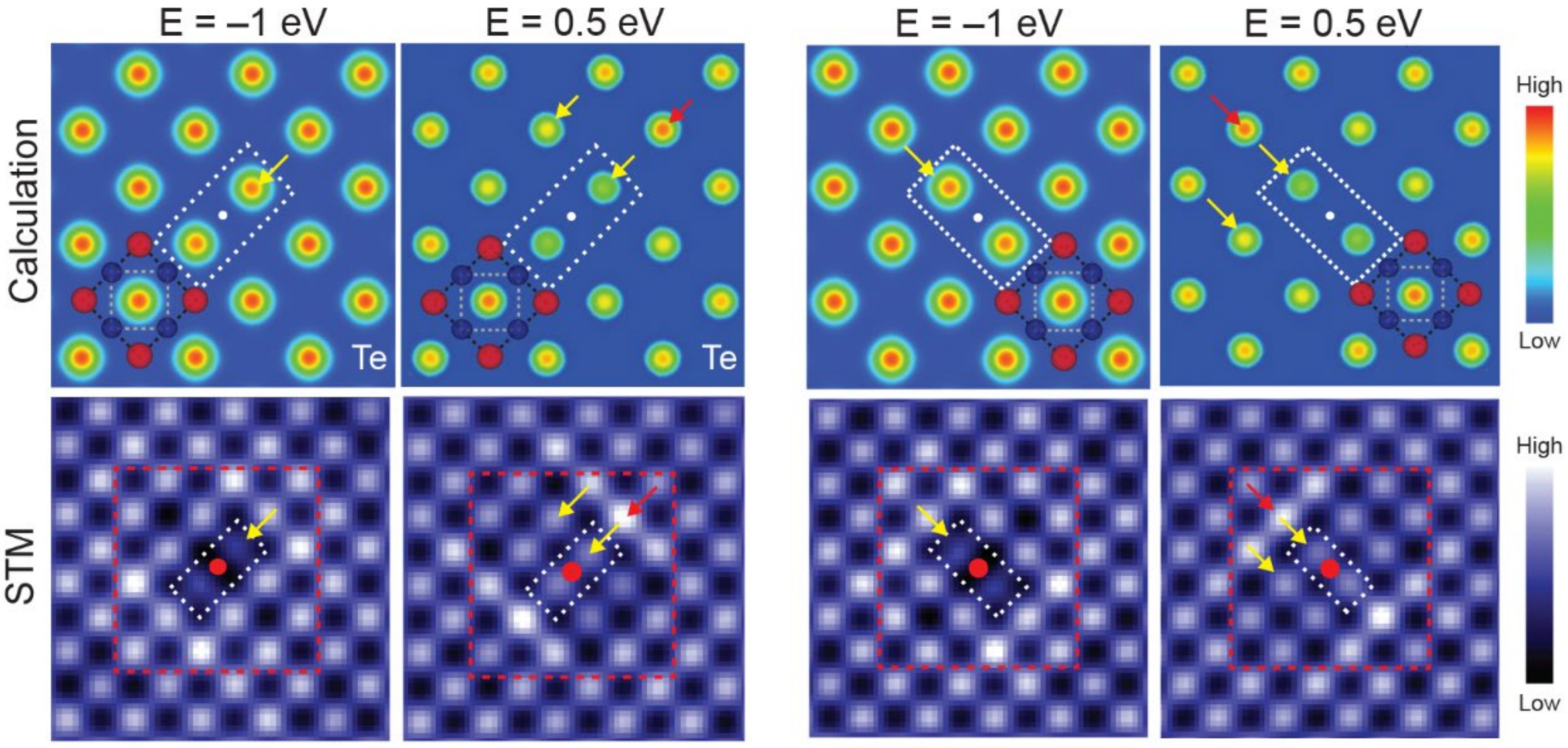

**Fig. S13.** Anisotropic impurity state induced by the Te-substitution to the Sb square-net layer.

(Top row) Simulated charge density distribution on the Te-surface along on two different $C_2$-symmetric Sb-sites as indicated in Fig. 1**a**. (Bottom row) Averaged topography over several $C_2$ defects from the same FOV in Fig 4**b**. The white rectangular boxes enclose two Te atoms in the immediate vicinity to the $Te_{Sb}$-sites below, which are marked by the white dots. The FOV corresponding to the simulation is represented in a dotted red box. The atoms with reduced and enhanced intensity due to $Te_{Sb}$-substitution are indicated by yellow and red arrows, respectively. Simulated images and averaged STM topography are in excellent agreement.

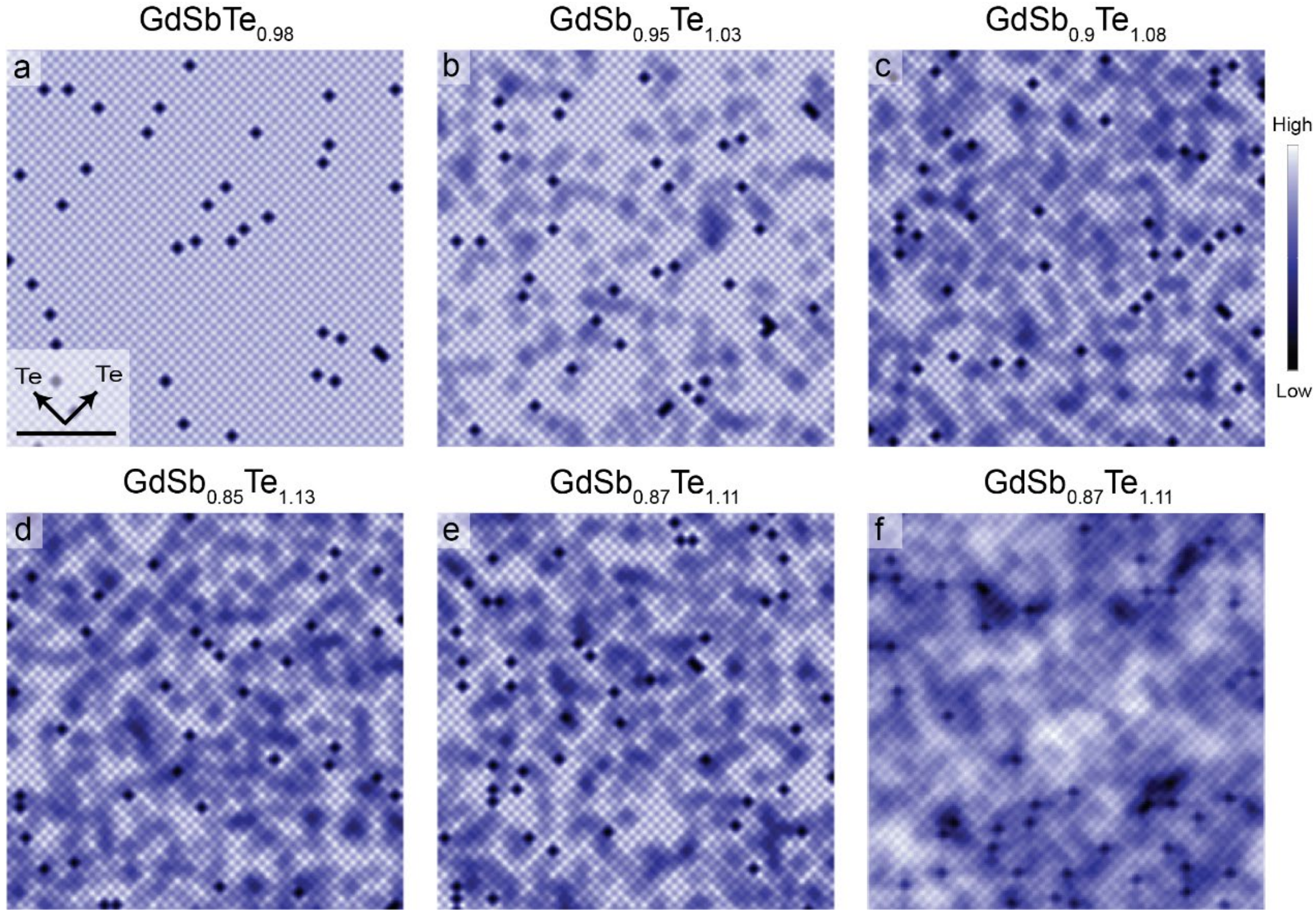

**Fig. S14.** Simulated topographic images of randomly distributed $C_2$–symmetric impurities.

**a** - **e** The simulated topography for the Te substitution in the Sb sites for the composition x = 0, 5, 10, 15 and 13% respectively with the 5nm scale bar. We use 2% Te vacancy in all simulated images. **f** The experimental data with the same doping in (**e**).

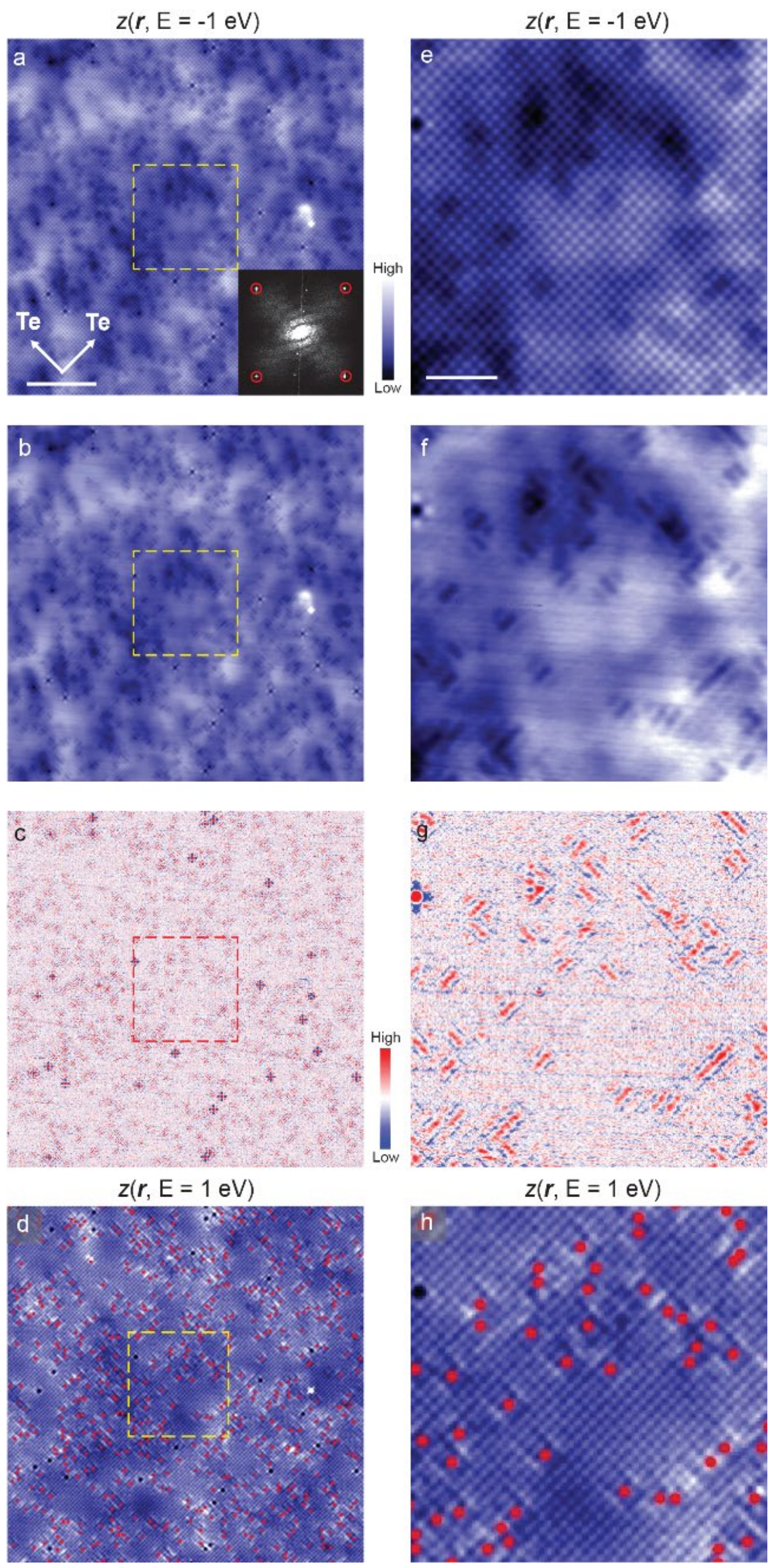

**Fig. S15.** Correlation between $C_2$ impurities and local unidirectional nanostructure in $GdSb_{0.98}Te_{1.02}$.

**a** Topography, $z(\boldsymbol{r}, E = -1\ \text{eV})$ acquired in the same FOV as in Fig. 4**b** with $V_{Bias} = -1V$ and $I = 20$ pA. The inset shows the corresponding Fourier transform with Bragg peaks marked by red circles. The scale bar represents 10 nm. **b** $z(\boldsymbol{r}, E = -1\ \text{eV})$ but after all Te atoms are Fourier-filtered from (**a**), showing the $C_2$-symmetric impurity states and background electronic inhomogeneity. **c** $z(\boldsymbol{r}, E = -1\ \text{eV})$ after we performed Laplacian on **b** to enhance the contrast of $C_2$-symmetric impurity states, allowing us to locate the Te-substituted Sb-sites. **d** Te-substituted Sb-sites, which are marked by red dots, overlapped on the topography, $z(\boldsymbol{r}, E = 1\ \text{eV})$ in Fig. 4**b**. **e – h** shows the enlarged images taken in the area marked by dashed square from (**a**) - (**d**), respectively. The scale bar in (**e**) represents 3 nm. The correlation between $C_2$ impurities and the local unidirectional nanostructure are evident in (**d**) and (**h**).

**Supplementary Note 2.** Charge modulation simulation

As in Supplementary Note 2, we use Eq. (1) to model the smectic charge modulation, $\boldsymbol{q_S}$. In the absence of disorder, the effective two-dimensional free energy is expended in terms of order parameters and their derivatives (*Ref. 21,22*):

$$F_{clean} = \int d^2r[C_1\left(|\partial_x\Phi_x|^2 + |\partial_y\Phi_y|^2\right) + C_2\left(|\partial_y\Phi_x|^2 + |\partial_x\Phi_y|^2\right) + s\left(|\Phi_x|^2 + |\Phi_y|^2\right) + \frac{u}{2}\left(|\Phi_x|^2 + |\Phi_y|^2\right)^2 + \nu|\Phi_x|^2|\Phi_y|^2] \quad \ldots\ldots (6)$$

The free energy respects all the symmetries of a square lattice, consistent with the fact that the lattice structure of $GbSb_{0.87}Te_{1.11}$ is tetragonal. The stability conditions require $u > 0$ for all phases and $2u + \nu > 0$ for the liquid phase. The coefficient $s = s(T)$ is a function of the temperature T. For T > $T_c$ (the critical temperature), $s > 0$, in accordance with a fully symmetric liquid phase. For T < Tc, $s < 0$, and non-zero order parameters give rise to broken symmetry phases. A broken symmetry phase can be either checkerboard ($\nu < 0$) or stripe ($\nu > 0$), depending on the sign of the coupling constant. The mean-field solution gives the phase diagram as shown in Fig. S16**a**.

To include the influence of quenched disorder, we consider an additional term of free energy:

$$F_{impurity} = -\int d^2r\left(H_x^*\Phi_x + H_y^*\Phi_y + c.c.\right) \ldots (7)$$

where $H_x$ and $H_y$ are two identical independent complex random fields, whose magnitudes are Gaussian random variables with mean zero and standard deviation $h$. Random phases are uniformly distributed on [0, 2π].

The total free energy is therefore:

$$F_{total} = F_{clean} + F_{impurity} \quad \ldots\ldots\ldots\ldots\ldots\ldots\ldots\ldots (8)$$

We discretize the total free energy to 142 × 142 lattice sites with periodic boundary conditions. In the presence of disorder, the result of numerically minimizing the total free energy suggests the proliferation of topological defects in broken symmetry phases. The individual components of the charge modulation pattern show electronic stripe characteristics. The resulting phase diagram is shown in Fig. S16, similar to that in *Ref. 21*.

Searching for low-energy states, we adopt the conjugate gradient method to minimize Eq. 8. The resulting total free energy density and field configuration are shown in Fig. S17. The vortex in $\Phi_x$ and $\Phi_y$ corresponds to the dislocations of the CDWs in this context. Vortices and anti-vortices usually come as a pair, such that the elastic energy relaxes at a large distance. The quenched disorder prevents vortex-antivortex pairs from recombining. Due to the highly frustrated nature of competition between elastic and disorder energy, different low-energy states are separated by a large barrier. Therefore, we expect (and observe) a long relaxation time. The topological defects are stable within the time scale of experimental measurements.

Using the definition of Eq. (1), the charge order is shown in Fig. S17**a** and the *x*-component and *y*-component of charge order are plotted in Fig. S18**b** and Fig. S18**c**, respectively, which successfully capture the dislocation features observed in our experimental study of $GbSb_{0.87}Te_{1.11}$. To extract the features of charge modulation from STM images, the experimental data undergo Fourier transformation to reciprocal space, exclude the spatial frequencies that are away from the wave vector of CDW, and then inverse Fourier transform the results to get the charge modulation pattern. The omitted high spatial frequencies inevitably cause blurs near dislocations.

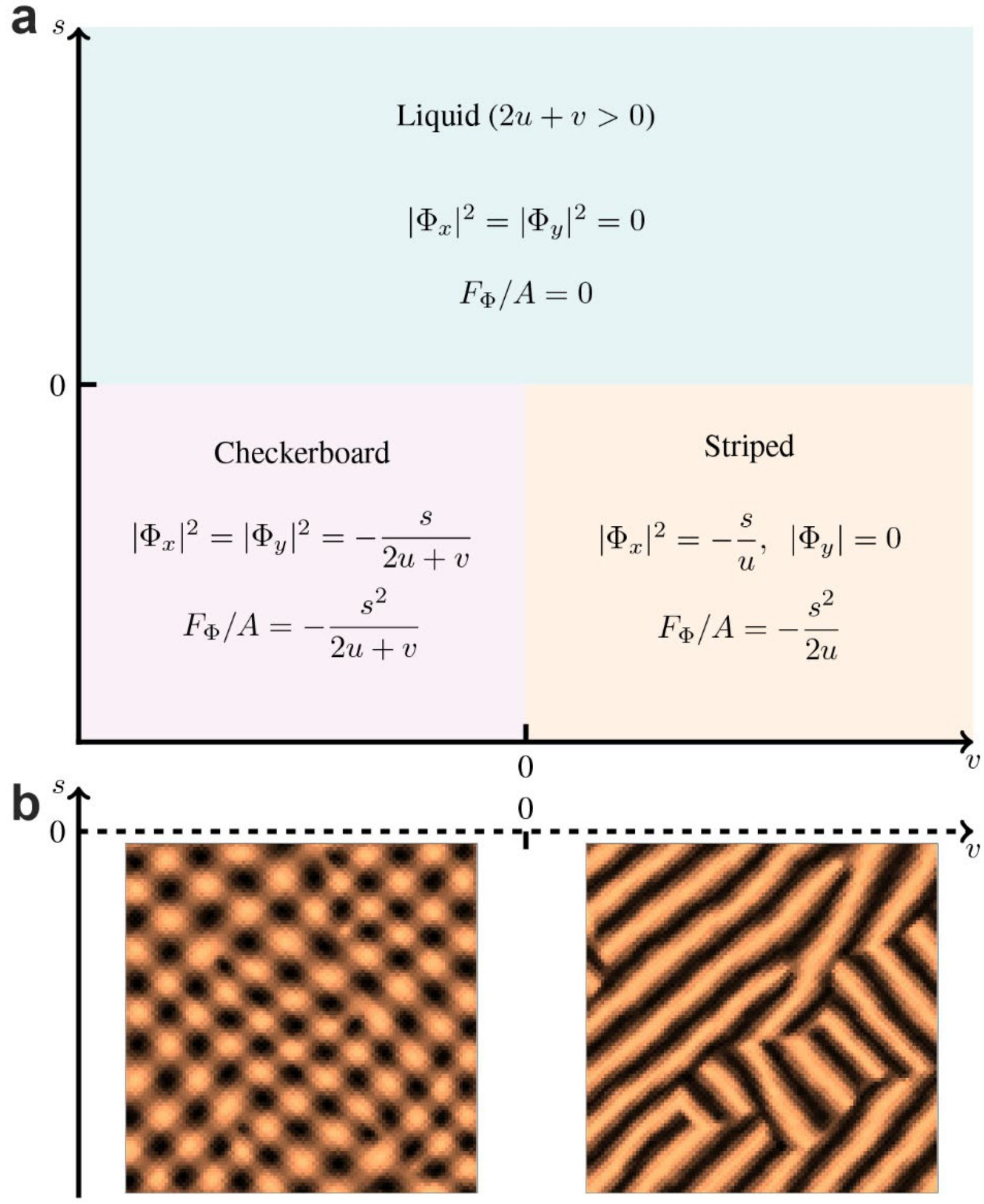

**Fig. S16.** Charge modulation simulation and correlation length analysis.

**a** The phase diagram is given by the mean-field solution of Eq. (6). **b** In the presence of quenched disorder, a broken symmetry phase becomes either a checkerboard with dislocations or a stripe with dislocations and domain walls. The dashed line indicates the transition between the symmetry liquid phase and the broken symmetry phase becomes crossover. Here *A* is the area of the unit cell.

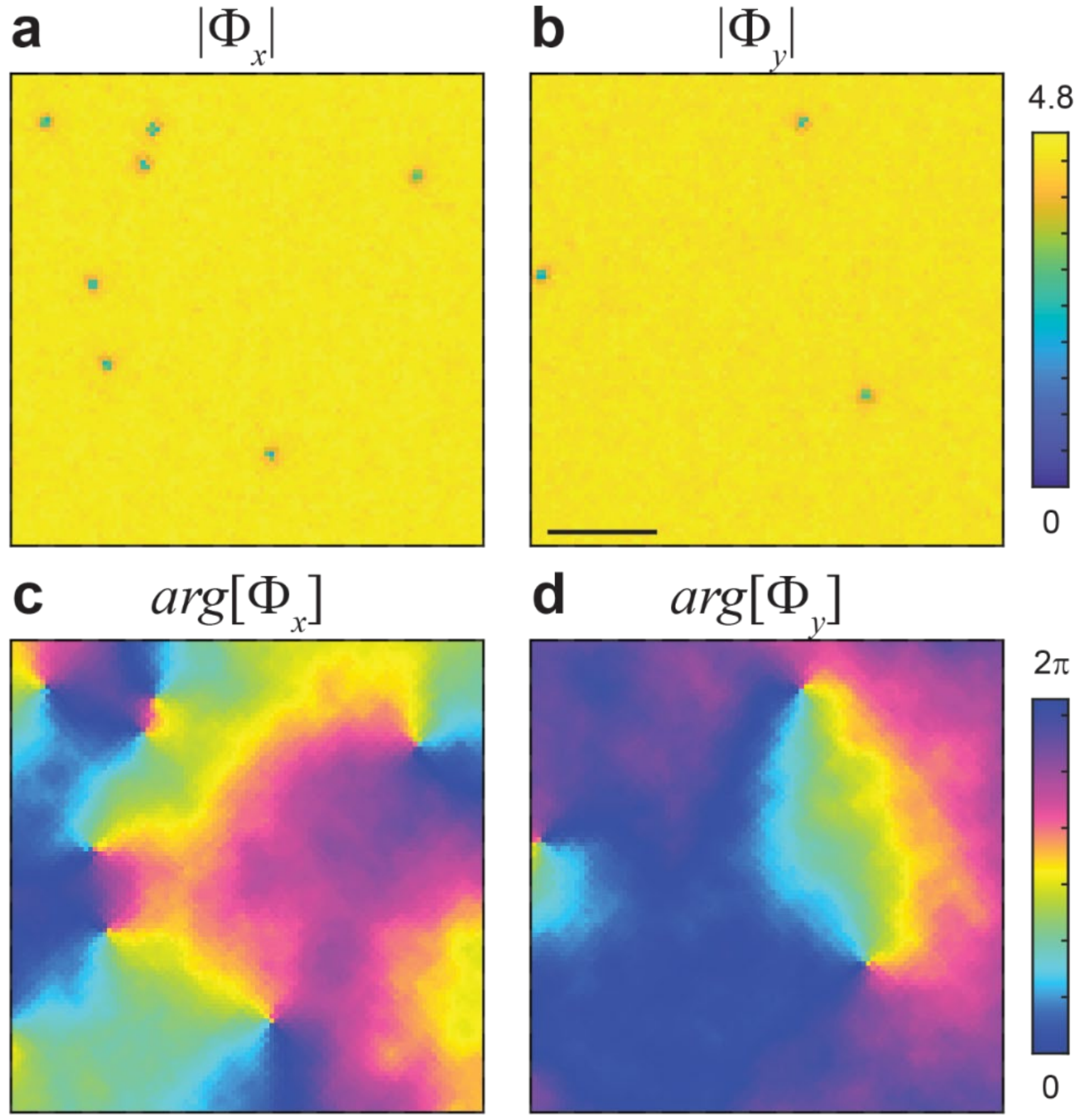

**Fig. S17.** The ground state field configuration.

We consider a disorder strength $h$ = 0.6 and then minimize the total free energy with parameters $C_1 = C_2 = 1$, $s = -0.1$, $u = 0.1$, $v = -0.1$, yielding the absolute value and the phases of complex fields $\Phi x$ and $\Phi y$ in **a** – **b** and in **c**– **d**, respectively. The scale bar represents 10 nm. The vortex structure can be clearly seen in (**c**) and (**d**).

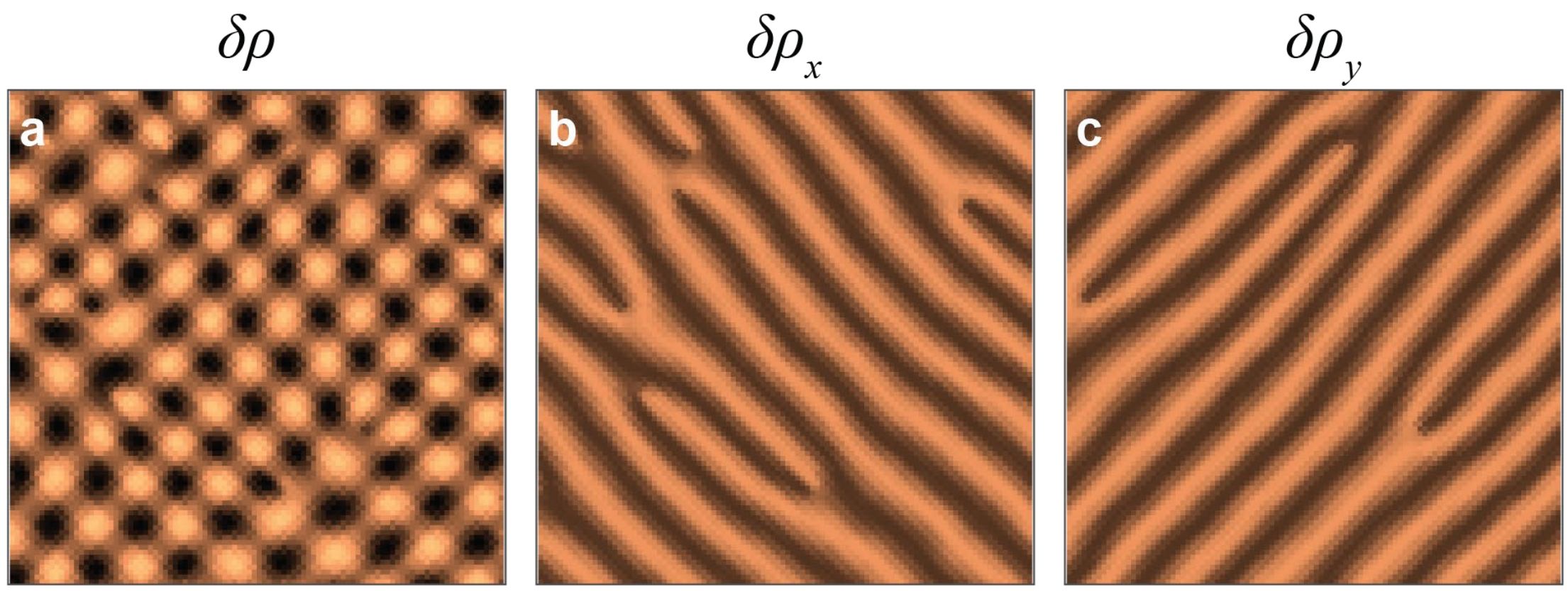

**Fig. S18.** Charge density fluctuations.

**a** Charge densities, which approximately corresponds to 43 nm × 43 nm in real space, results from the same field configuration as in Fig. S17 using Eq. (1). **b** and **c** $x$-component and $y$-component of charge density decomposed from (**a**), respectively.

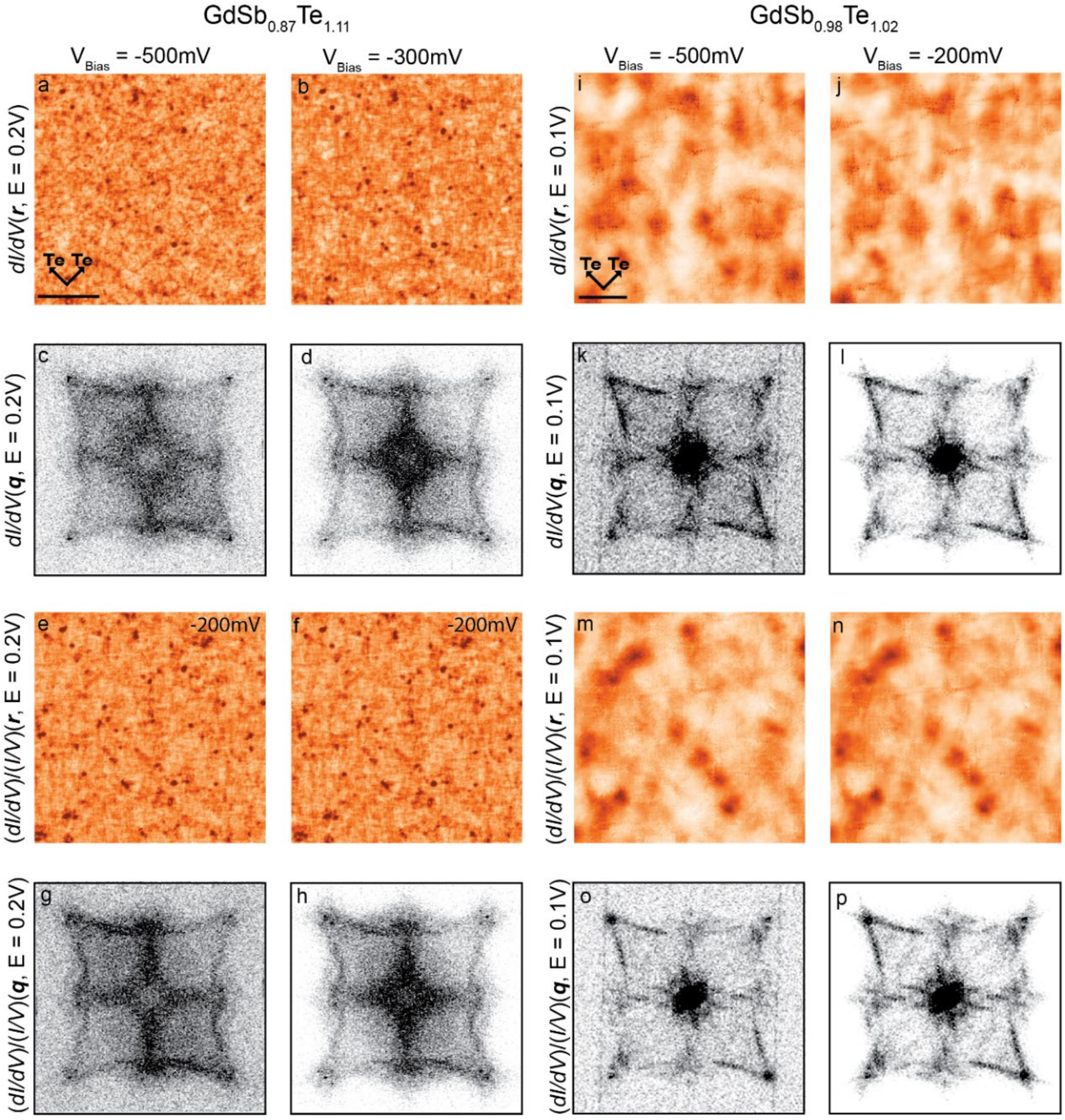

**Fig. S19.** Setpoint effect on the conductance maps.

**a** and **b** Differential conductance map for $GdSb_{0.87}Te_{1.11}$ at E = -0.2 eV with $V_{bias}$ = -500 mV, -300 mV respectively. **c** and **d** are their corresponding Fourier transformed images, *dI/dV*(***q***, E). **e** and **f** Feenstra normalized differential conductance map at E = -0.2eV with $V_{bias}$ = -500 mVand -300 mV, respectively. **g** and **h** are the Fourier transform of **e** and **f**, respectively. **i** and **j** Differential conductance map for $GdSb_{0.98}Te_{1.02}$ at E = -0.1eV with $V_{bias}$ = -500 mV and -200 mV, respectively. **k** and **l** are their corresponding Fourier transformed images, *dI/dV*(***q***, E). **m** and **n** are the Feenstra normalized differential conductance maps of **i** and **j**. **o** and **p** are the Fourier transform images of **m** and **n**, respectively.

**Table S1.** Sommerfeld coefficient, $\gamma$ of *Ln*SbTe.
For comparison, $\gamma$ of ZrSiS is 6.84 mJ/mol-K$^2$, which is adapted from *Ref.* 20.

| ***Ln*SbTe** | **$\gamma$ (mJ/mol-K$^2$)** | **Reference** |
|---|---|---|
| LaSbTe | 0.51 | K. Pandey *et al.*, Crystals 12, 1663 (2022). |
| CeSbTe | 41 | B. Lv *et al.*, J. Phys.: Condens. Matter 31 355601 (2019) |
| PrSbTe | 2.6 | D. Yuan *et al.*, Phys. Rev. B 109, 045113 (2024) |
| NdSbTe | 115 | K. Pandey *et al.*, Phys. Rev. B 101, 235161 (2020) |
| PmSbTe | N/A | |
| SmSbTe | 160 | K. Pandey *et al.*, Adv. Quantum Technol. 4, 2100063 (2021). |
| EuSbTe | N/A | |
| GdSbTe | 7.6 | R. Sankar *et al.*, Inorg. Chem. 58, 11730 (2019) |
| TbSbTe | 0.66 | Fei Gao *et al.*, Adv. Quantum Technol. 6, 2200163 (2023) |
| DySbTe | 0.45 | Fei Gao *et al.*, Phys. Rev. B 105, 214434 (2022) |
| HoSbTe | 382 | Meng Yang *et al.*, Phys. Rev. Mater. 4, 094203 (2020) |
| ErSbTe | N/A | |